\documentclass[%
 reprint,
amsmath, amssymb,
aps,
]{revtex4-2}
\usepackage{graphicx}
\usepackage{makecell}
\usepackage{dcolumn}% Align table columns on the decimal point
\usepackage{bm}% bold math
\usepackage{xcolor}
\usepackage{graphicx}
\usepackage{epsfig}
\usepackage{slashed}
\usepackage{tabu,multirow}
\usepackage[unicode=true,bookmarks=false,breaklinks=false,pdfborder={0 0 1},colorlinks=true]
 {hyperref}
\hypersetup{
 citecolor=blue,linkcolor=blue,urlcolor=blue}

\begin{document}

\preprint{APS/123-QED}

\title{Holographic stability for non-$q\bar{q}$ candidates}% Force line breaks with \\

\author{Miguel Angel Martin Contreras}
\email{miguelangel.martin@usc.edu.cn}
 %\affiliation{%
%  Instituto de F\'isica y Astronom\'ia, \\
%  Universidad de Valpara\'iso,\\
%  A. Gran Breta\~na 1111, Valpara\'iso, Chile}
\affiliation{
 School of Nuclear Science  and Technology\\
 University of South China\\
 Hengyang, China\\
 No 28, West Changsheng Road, Hengyang City, Hunan Province, China.
}

\author{Alfredo Vega}%
 \email{alfredo.vega@uv.cl}
\affiliation{%
 Instituto de F\'isica y Astronom\'ia, \\
 Universidad de Valpara\'iso,\\
 A. Gran Breta\~na 1111, Valpara\'iso, Chile
}

% \collaboration{MUSO Collaboration}%\noaffiliation

%  \homepage{http://www.Second.institution.edu/~Charlie.Author}
% \affiliation{
%  Second institution and address\\
%  This line break forced% with \\
% }%
% \affiliation{
%  Third institution, the second for Charlie Author
% \affiliation{%
%  Authors' institution and address\\
%  This line break forced with \textbackslash\textbackslash
% }%


\date{\today}% It is always \today, today,

\begin{abstract}
In the context of bottom-up AdS/QCD models, we discuss how the configurational entropy can describe heavy non-$q\,\bar{q}$ states. Using the nonquadratic softwall model, introduced to describe nonlinear Regge trajectories, we parametrize different multiquark and exotic meson structures to describe $Z_c$, $\psi$, and $Z_b$ states as non-$q\,\bar{q}$ hadrons in terms of stability. We found that $Z_c$ is better described as a hybrid meson with one gluon tube,  $\psi$ as hadrocharmonium, and $Z_b$ as hadronic molecule.
% \begin{description}
% Secondary publications and information retrieval purposes.
% You may use the \texttt{description} environment to structure your abstract;
% Use the optional argument of the \verb+\item+ command to give the category of each item. 
% \end{description}
\end{abstract}

\maketitle

%\tableofcontents

\section{Introduction}
Since the seminal work \cite{Karch:2006pv}, the softwall model has become one of the most famous bottom-up models in AdS/QCD to describe a plethora of hadronic properties such as form factors \cite{Grigoryan:2007my, Abidin:2009hr}, deconfinement phase transition \cite{Herzog:2006ra, Cai:2007zw, Afonin:2018era}, chiral symmetry breaking \cite{, Ballon-Bayona:2021ibm}, scattering processes \cite{BallonBayona:2007qr, Braga:2011wa, Vega:2010ns}, and hadron spectroscopy \cite{Colangelo:2008us, Vega:2008af, Vega:2008te, Branz:2010ub, Gutsche:2012wb}. This model has also been extended to dynamical versions \cite{Batell:2008zm, Li:2012ay, Li:2013oda}, finite density \cite{Colangelo:2010pe, Sachan:2011iy}, finite temperature \cite{Kim:2007rt, Fujita:2009ca, Fujita:2009wc}, and magnetic field \cite{Dudal:2015wfn, Li:2016gfn, Fang:2021ucy}. 

The main idea behind this model is the emergence of bounded states in the AdS bulk because of a dilaton field that breaks conformal invariance. These bounded states are dual to hadrons at the boundary. The matching is achieved by comparing the two-point function calculated in large $N$ QCD at high $q^2$ with the holographic dual object. The most particular feature of the softwall model is the linear confinement, i.e., the bounded states form a linear mass spectrum, dual to Regge trajectories.  These trajectories are suitable for describing light unflavored mesons. However, when the constituent masses start to be considered, the linearity in the trajectories disappears \cite{Chen:2018nnr}.

An alternative to describe these nonlinearities in hadrons spectra is a deformation of the quadratic structure in the dilaton field, inherited from the softwall model. This idea was applied to describe isovector mesons in \cite{MartinContreras:2020cyg}. Including the constituent mass makes it possible to test other hadronic species as non-$q\,\bar{q}$ states. In particular, this work wants to explore the configurational entropy as a test tool to describe these states, as it was initially exposed in \cite{Colangelo:2018mrt, Karapetyan:2021ufz}. In the first work \cite{Colangelo:2008us}, the authors explore how configurational entropy (CE) is related to multiquark hybrid meson states with high CE value.
On the other hand, \cite{Karapetyan:2021ufz} discusses how CE is a tool for describing the abundance of these non-$q\,\bar{q}$ states in nature. This work follows a different approach. We will connect the configurational entropy with the hadron stability by discussing how bulk locality is equivalent to confinement at the conformal boundary. This will motivate the possibility of using CE to distinguish what multiquark or hybrid meson models are suitable for describing heavy quark exotica. 

This manuscript is organized as follows: In Sec. \ref{non-q}, we give a summary of the ideas behind the nonquadratic softwall model introduced in \cite{MartinContreras:2020cyg}, and how it works describing non-$q\, \bar{q}$ states. Section \ref{entropy} describes the differential configurational entropy (DCE) algorithm. Section \ref{hadron-stability} discusses the connection between DCE and hadron stability. The application of DCE to what non-$q\,\bar{q}$ structures are suitable to describe heavy vector exotica is written in Sec. \ref{DCE-hadron}. Finally, we summarized our work in Sec.  \ref{conclusions}. 

\section{Nonquadratic softwall model}\label{non-q}
\subsection{Nonquadratic dilaton in a nutshell}
The nonquadratic dilaton is a holographic proposal developed to address heavy quarkonium masses with better accuracy \cite{MartinContreras:2020cyg}. The motivation comes from the Bethe-Salpeter formalism, where Regge trajectories, when including constituent quark masses,  can be written as \cite{Chen:2018bbr}

\begin{equation}
(M_n-m_{q_1}-m_{q_2})^2=a(n+b),    
\end{equation}

\noindent where $a$ is a universal slope and $b$ accounts for the effect of the mesonic quantum numbers, it is expected that nonlinearities associated with the constituent mass emerge \cite{Afonin:2014nya, Chen:2018hnx}. In the holographic AdS/QCD context, it is possible to include these constituent quark mass effects by adding an extra $\nu$ exponent to the radial trajectory as follows:

\begin{equation}\label{nonlinear-fit}
M_n^2=a(n+b)^\nu.    
\end{equation}

This nonlinearity in Regge trajectories is translated into deformations on the static quadratic dilaton 

\begin{equation}
    \Phi(z)=\left(\kappa\,z\right)^{2-\alpha},
\end{equation}

\noindent where the SWM dilaton gains an extra parameter $\alpha$ that captures the constituent mass effects.  In this sense, $\kappa$ and $\alpha$  set a running with the constituent mass of hadrons in a particular family. In the original work \cite{MartinContreras:2020cyg}, the isovector mesons,-are composed of $\rho$, $\omega$, $J/\psi$, and $\Upsilon$ mesons, and labeled $I^G\,J^{PC}=0^-(1^{--})$. The $\rho$ meson corresponds to the light unflavored case, where mesons, under chiral symmetry, have zero constituent mass. This mass scheme sets the following parametrizations for $\kappa$ and $\alpha$

\begin{eqnarray}\label{kappa-1}
\alpha(\bar{m}) &=& a_{\alpha}-b_{\alpha} e^{-c_{\alpha} \bar{m}^2}\\ \label{alpha-1}
\kappa (\bar{m}) &=& a_{\kappa} - b_{\kappa}e^{-c_{\kappa} \bar{m}^{2}},
\end{eqnarray}

\noindent where the fit coefficients are given by 

\begin{equation*}
    \begin{gathered}
    a_{\alpha} = 0.8454,~b_{\alpha} = 0.8485,~c_{\alpha} = 0.4233\,\, \text{GeV}^2,\\
a_{\kappa} = 15.2085\,\text{GeV},~b_{\kappa} = 14.8082\,\text{GeV},~c_{\kappa} = 0.0524\,\text{GeV}^2
    \end{gathered}
\end{equation*}

From the running of $\kappa$ and $\alpha$ with the hadron constituent mass, i.e., $\kappa(\bar{m})$ and $\alpha(\bar{m})$, we can infer specific values for $\kappa$ and $\alpha$ for a given vector hadron, as a non-$q\,\bar{q}$ candidate. In this procedure, the input is the parameterization of the constituent mass $\bar{m}$, which contains information on the inner structure of the hadron. Thus, we will fix the constituent mass parametrization instead of setting a Regge scale and a quadratic deformation exponent. 

\subsection{Holographic setup}
As customary in these models,  the starting point is a density action for the bulk field dual to vector hadrons:

\begin{equation}
    I_\text{H}=-\frac{1}{2}\int{d^5x\,\sqrt{-g}\,e^{-\phi(z)}\,\left(\frac{1}{2\,g^2_5}\,F_{mn}F^{mn}-M_5^2\,A_m\,A^m\right)},
\end{equation}

\noindent {where $g_5^2=\frac{12\,\pi^2}{N_c}$ is fixed from the large $q^2$ expansion of the two-point function at the boundary \cite{Erlich:2005qh}.

This action lives in an AdS$_5$ space described by the Poincarè patch:

\begin{equation}
    dS^2=\frac{R^2}{z^2}\left(dz^2+\eta_{\mu\,\nu}\,dx^\mu\,dx^\nu\right), 
\end{equation}

\noindent where $R$ is the AdS  curvature radius and $\eta_{\mu\nu}=\text{diag}(-1,\vec{1})$. 

At this point, mentioning the gauge invariance is essential since we now have massive vector bulk fields. Recall that the gauge invariance should be manifest at the conformal boundary, where all dual fields are massless \cite{Braga:2015lck}. The nonzero bulk mass does not affect the gauge $A_z=0$, which is imposed to decouple the bulk from the boundary information. If we pay attention to the massive e.o.m for the vector bulk fields, i.e., for the $z$ component

\begin{equation}
\Box\,A_z-\partial_z\left(\partial_\mu\,A^\mu\right)+M_5^2\,e^{2\,A}\,A_z=0,    
\end{equation}

\noindent imposing the $A_z=0$ gauge still implies the transverse gauge $\partial_\mu\,A^\mu=0$ at the conformal boundary, so Ward-Takahashi identities hold. 

From this configuration, after writing the equations of motion in the $A_z=0$ gauge, imposing the on-shell mass condition $M_n^2=-q^2$,  Fourier transforming the massive bulk field $A_\mu(z,q)=\tilde{A}_\mu(q)\,\psi_{z,q}$ and performing the standard Boguliubov transformation $\psi(z)=e^{B(z)/2}\phi(z)$\cite{Karch:2006pv}, we can reduce the hadron spectroscopy problem to solve a Schrödinger-like equation

\begin{equation}\label{Schrodinger-equ}
    -\phi''(z)+V(z)\phi(z)=M_n^2\,\phi(z),
\end{equation}

\noindent where the holographic potential $V(z)$ is defined in terms of the Bogoliubov transformation $B(z)$ function  as

\begin{equation}\label{holo-pot}
    V(z)=\frac{B'(z)^2}{4}-\frac{B''(z)}{2}+\frac{M_5(\Delta)^2\,R^2}{z^2}
\end{equation}

\noindent with $B(z)=\Phi(z)-\log(R/z)$. This $B$ function is fundamental in the bottom-up AdS/QCD scenario since it encloses how the background captures hadronic physics at the boundary. The other ingredient is the bulk mass $M_5$.

From the field/operator duality, the operator creating hadrons at the boundary is dual to the bulk field. The most general operator that creates hadrons is a function of the constituent fields (quarks and gluons) and the covariant derivatives, that is, $\mathcal{O} = f(\bar{q},q, G_{\mu\,\nu}, D_\mu)$. This operator definition allows us to consider more general structures, such as \emph{hadronic molecules} or \emph{hadroquakonium} in the non-$q\,\bar{q}$ phenomenology.

The dimension of $\mathcal{O}$, $\text{dim}\,\mathcal{O}=\Delta$ controls how the bulk normalizable mode behaves near the conformal boundary, i.e., $\psi(z\to0)\to z^{\Delta-1}$ for vector fields. Imposing that the bulk field spin is equivalent to the hadronic spin $S$, we can write the most general scaling dimension $\Delta$ in terms of hadronic information \cite{Vega:2008te}

\begin{equation}
\Delta =\Delta_0+L+\text{anomalous dimensions}, 
\end{equation}

\noindent where $\Delta_0$ is associated with pure constituent information and $L$, counting covariant derivatives, sets the hadronic angular momentum number. In the most general situation, including data from the renormalization group is possible. This information can be encoded in these anomalous dimensions terms \cite{Gursoy:2007cb, Cherman:2008eh, Boschi-Filho:2012ijd}.  For simplicity, we will not include anomalous dimensions. 

This relation for the scaling dimension, via the equations of motion for the vector field, allows us to write an expression for the hadronic mass identity, encoded into the bulk mass as follows

\begin{equation}
    M_5^2\,R^2=\left(\Delta_0+L-1\right)\left(\Delta_0+L-3\right).
\end{equation}

However, this expression for vector hadrons does not unambiguously define vector hadrons. Consider non-$q\,\bar{q}$ structures with spin one as diquark-antidiquark pairs, hadronic molecules, and hadroquarkonium. For a given $L$, all of these tetraquark structures have $\Delta_0=6$. Thus, they have the same bulk mass, i.e., $M_5^2\, R^2=15$. This \emph{degeneracy} can be lifted in the nonquadratic dilaton scenario since $\kappa$ and $\alpha$ are sensible to the hadron structure. 

Once the hadronic identity is defined, we can focus on the holographic potential for the  non-$q\,\bar{q}$ vector states. In the bottom-up context, once we have defined the dilaton field, i.e., $\Phi(z)=\left(\kappa\,z\right)^{2-\alpha}$, it is straightforward to write down the holographic potential using the expression \eqref{holo-pot} and the Boguliubov transformation

\begin{equation}
    B(z)=\left(\kappa\,z\right)^{2-\alpha}-\log\left(\frac{R}{z}\right).
\end{equation}

Therefore, the holographic potential for non-$q\bar{q}$ states acquires the following structure:
%In this nonquadratic model, \textcolor{red}{using the expression \eqref{holo-pot}}, has the following form 

 \begin{equation} \label{non-qqbar-pot}
V_{\text{non-}q\,\bar{q}}(z,\kappa,\,\Delta)=V_{q\,\bar{q}}(z,\kappa,\alpha)+\frac{M_5^2(\Delta)\,R^2}{z^2},      
 \end{equation}

\noindent where $V_{q\,\bar{q}}(z)$ is the potential used to compute the masses of the isovector family in the $s$-wave. It is defined as 

\begin{multline}\label{nonlinear-pot}
V_{q\,\bar{q}}(z,\kappa,\alpha)=\frac{3}{4 z^2}-\frac{1}{2} \alpha ^2 \,\kappa^2 \,(\kappa\,  z)^{-\alpha }+\frac{1}{4} \alpha^2\, \kappa^2 (\kappa\,  z)^{2-2\alpha }\\
+\frac{3}{2} \alpha \, \kappa ^2 \,(\kappa \, z)^{-\alpha }-\kappa ^2 (\kappa \, z)^{-\alpha }-\alpha \, \kappa ^2\, (\kappa \,z)^{2-2 \alpha }\\
+\kappa^2\, (\kappa \, z)^{2-2 \alpha }+\frac{\kappa }{z}(\kappa\,  z)^{1-\alpha }-\frac{\alpha\,\kappa  }{2\, z} (\kappa\,  z)^{1-\alpha}.   
\end{multline}

Setting $\alpha=0$ in $V_{q\,\bar{q}}(z)$ reduces the potential to the well-known softwall model expression \cite{Karch:2006pv}. 

The holographic algorithm will now focus on defining a threshold mass $\bar{m}$, capturing the information about the structure of hadrons in terms of their constituents. This threshold mass will define a pair $\kappa(\bar{m})$  and $\alpha(\bar{m})$ for each structure. We can compute the hadronic squared mass from the holographic potentials with these inputs. With the eigenmodes, it is straightforward to compute the associated configurational entropy. 

In the next sections, we will follow the following recipe: once we define $\kappa$ and $\alpha$ using a proper threshold mass $\bar{m}$, we set them as entries into the holographic potential \eqref{non-qqbar-pot} to solve the eigenvalue problem \eqref{Schrodinger-equ}.

\subsection{Holographic non-$q\,\bar{q}$ candidates}
\emph{Exotic hadrons} are 
all the mesonic states with quantum numbers not allowed by the usual $q\,\bar{q}$ model. A good review of the physics of such states can be found in  \cite{Brambilla:2019esw, Guo:2017jvc, Lebed:2016hpi} and references therein. In top-down AdS/QCD,  the work \cite{PhysRevD.100.126023} addresses the exotic meson spectra for $Z_c$ and $Z_b$ in the context of Sakai-Sugimoto models.

This section will focus on three candidates for tetraquarks: $Z_c$, $\psi$, and $Z_b$ states. We will test diquark-antidiquark, hadroquarkonium, hadronic molecule, and hybrid meson configurations. The first three configurations, known as \emph{multiquark states}, have $\Delta=6$. The last configuration, known as \emph{gluonic excitations}, has $\Delta=5$ for a single gluonic tube and $\Delta=7$ for two gluon tubes. All of these states are considered in the $s$-wave. 

To give a parametric description of the hadronic constituent mass threshold $\bar{m}_{eh}$, we will consider a generic exotic hadron with $N$ constituents that can be quarks, gluons, or mesons. Each of these constituents will contribute in a \emph{weighted} form to $\bar{m}_{eh}$ as \cite{MartinContreras:2020cyg}

\begin{equation}\label{exotic-mass-conf}
\bar{m}_{eh} =\sum_{i=1}^N({P^\text{quark}_i\,\bar{m}_{q_i}+P^\text{Gluon}_i\,m_{G_i}+P^\text{meson}_i\,m_{\text{meson}_i}}),   
\end{equation}

\noindent along with the constraint condition

\begin{equation}\label{constrain-hadron}
\sum_{i=1}^N\,(P^\text{quark}_i+P^\text{Gluon}_i+P^\text{meson}_i)=1.    
\end{equation}

Notice that each weight $P_i$ measures the contribution of a given constituent (quark, gluon, or meson) having a defined constituent mass. In our case, we choose the following constituent masses 

\begin{equation*}
    \begin{gathered}
m_{u} = 0.336~\text{GeV},~m_{d}=0.340~\text{GeV}\\
m_{c} = 1.550~\text{GeV},~m_{b}=4.730~\text{GeV}\\
m_G=0.7~\text{GeV},~m_{\rho(770)}=0.775~\text{GeV}\\
m_{J/\psi}=3.097~\text{GeV},~ m_{\Upsilon(1S)}=9.460~\text{GeV}.
    \end{gathered}
\end{equation*}

Mesons masses are extracted from \cite{Workman:2022ynf}. Gluon constituent mass is read from \cite{Hou:2001ig}.

The threshold mass for exotic hadrons is an extension of the proposal done in Ref. \cite{MartinContreras:2020cyg}. Since bottom-up models do not have \emph{ab initio} information about the hadron inner structure in the holographic dictionary, the threshold mass $\bar{m}$ opens the possibility of including it from pure phenomenological grounds. For the mesons, such as the isovector or heavy-light ones, the arithmetic average seems a reasonable choice since the dilaton slope $\kappa$, which fixes the scale for the hadronic Regge trajectories, would carry information about the constituent mass. This affirmation is not new in holography. For example, in top/down approaches, such as the D3/D7 systems, the quark mass plays a double role: it sets the embedding distance between the D3 and D7 branes, and it also sets the energy scale for the Regge trajectories that are not linear with the excitation number, i.e. $M_n^2\propto m_q^2\, n^2$ \cite{Kruczenski:2003be}.

By the same token, following the phenomenology of these non-$q\bar{q}$ candidates, it is possible to infer the inner structure from the decay modes. Following \cite{Brambilla:2019esw,Guo:2017jvc}, hadroquarkonium states have heavy meson cores appearing in the decay modes along with pion. For example, for the hadrocharmonium candidate, the $\psi(4230)$ state is expected to measure a $J/\psi\,\pi$  mode. Thus, expecting to have a larger contribution from a $J/\psi$ core in the threshold mass is consistent. A similar situation occurs with the hadronic molecule, hybrid mesons, and diquark pairs: the threshold mass is motivated by the decay modes expected for these states.

%%%%%%
%\subsubsection{Multiquark states}
%\textcolor{red}{Check quantum numbers at PDG}
\subsubsection{Diquark configuration}
 \emph{Diquarks} are noncolored singlet objects used as essential building blocks forming tetraquark mesons and pentaquark baryons. These fundamental blocks are either a color antitriplet or a color sextet in the SU(3) color representation \cite{Jaffe:2004ph}. Spin-spin interactions bound these diquarks. The constituent diquark approach helps describe the spectroscopy and decay of multiquark states. These diquark-composed candidates are expected to appear as poles in the $S$-matrix, described by narrow widths. 

Experimentally, charmonium and bottomonium tetraquark states can be identified thanks to their decay into open-flavor states instead of a quarkonium with a light meson due to the \emph{spin-spin interaction dominance} (See \cite{Brambilla:2019esw}). 

In the case of charmonium, $Z_c$ states, with quantum numbers $I^G(J^{CP})=1^+(1^{+-})$, are candidates to be vector tetraquarks. Following Bambrilla, we consider the $Z_c$ states as a single trajectory.  Other studies, such as \cite{Wang:2018ntv}, suggest $\psi(4230)$ with $0^+(1^{--})$ as a vector tetraquark instead of $Z_c(4200)$. However, in this holographic picture, considering both diquark-antidiquark structures implies the same threshold mass, $\bar{m}_{4q}$, leading to degeneration in mass. 

Holographically, a hadron is conceived as a bag with $N$ constituents characterized by $M_5$ despite its inner configuration. The holographic approach to the tetraquark is equivalent to having four quarks with $\Delta=3/2$, implying for the tetraquark $M_5^2\, R^2=15$. 

For the threshold mass $\bar{m}_{4q}$, without any loss of generality, we can assume that each $c$ quark contributes equally, i.e., $P_c=0.25$. Thus, we choose $\bar{m}_{4q}=m_c=1.55$ GeV as the threshold mass, implying that $\kappa_{4q}(\bar{m})=2.151$ GeV and $\alpha_{4q}(\bar{m})=0.5387$, see Table \ref{tab:one}. 

\subsubsection{Hadroquarkonium}
\emph{Hadroquarkonium states} are structures considering a vector meson core with a \emph{cloud} of two quarks \cite{Liu:2019zoy}. Experiments showed that most of the candidates to be heavy exotic states appear as final states composed of heavy quarkonium and light quarks,  motivating the idea that these states are a compact heavy quarkonium core surrounded by a light quark cloud \cite{Voloshin:2007dx}.  This quarkonium core interacts with the light quark cloud through a \emph{colored Van der Waals force}, allowing the decay of these states into the observed quarkonium core and the light quarks \cite{Brambilla:2017ffe}.  

At the holographic level, hadroquarkonium structures have the same behavior as tetraquarks, i.e., $\Delta=6$ and $M_5^2\, R^2=15$, since essentially, both models are a bag with four constituent quarks. The threshold mass settles the difference. Thus, for the hadrocharmonium threshold mass, we have

\begin{eqnarray*}
\bar{m}_{HQc}&=&\frac{1}{2}m_{J/\psi}+\frac{1}{4}\,(\bar{m}_u+\bar{m}_d)\\
&=& 1.717~\text{GeV}.
\end{eqnarray*}

%Thus, we find that $\kappa_{HC}(\bar{m}_{HQ}) 2.523$ GeV and $\alpha_{HC}(\bar{m}_{HQc})=0.604$.
We will consider that states  $\psi(4260)$, $\psi(4360)$ and $\psi(4660)$ with $0^+(1^{--})$ are hadrocharmonium  candidates \cite{Brambilla:2017ffe}, forming a vector trajectory. 

The summary of the holographic spectrum for $\psi$ states considered as hadrocharmonium is given in Table \ref{tab:two}. 

\subsubsection{Hadronic Molecules}
 \emph{Hadronic molecules} are states conformed by a pair of internal mesons bounded by strong QCD forces, interacting between them via a residual weak QCD colorless force \cite{Lebed:2016hpi}. These structures usually have two heavy quarkonia interacting or one heavy quarkonium plus a light meson. 

Recent works \cite{Guo:2017jvc, Brambilla:2019esw} suggest that the $Z_c$ or $\psi$ states are candidates for being charmonium hadronic molecules, having at least one pair of $c\,\bar{c}$ in the inner core of the molecule. These states usually decay to $J/\psi\,\pi$. 

On the holographic side, hadronic molecules have the same $\Delta$ and $M_5$ as tetraquarks and hadroquarkonium since the constituent content is essentially the same. Degeneracy is broken down by the threshold mass $\bar{m}_{HM}$ that captures information about the inner structure. For these charmonium hadronic molecules, we have 

\begin{eqnarray*}
 \bar{m}_{HM,\,\psi}&=&\frac{1}{3}m_{J/\psi}+\frac{2} {3}m_\rho=1.549\,\text{GeV},\\  \bar{m}_{HM,\,Z_c}&=&0.283 \,m_{J/\psi}+0.717\,m_\rho=1.4323\,\text{GeV}.\\    
\end{eqnarray*}

A similar situation occurs in the bottomonium sector \cite{Guo:2017jvc}, where $Z_B$ states, with $I^G(J^{CP})=1^+(1^{+-})$, are considered candidates for hadronic molecules, with a $\Upsilon(1\, S)$ meson core. In this situation, the threshold mass is 

\begin{equation}
\bar{m}_{HM,\,Z_B}=0.458\,m_{\Upsilon(1S)} +0.542\, m_\rho=4.753\,\text{GeV}. 
\end{equation}

%A summary of the holographic spectra associated with these hadronic molecule structures is given in Table XXXX.

The summary of the holographic spectra for the $Z_c$, $\psi$, and $Z_c$ states, considered hadronic molecules, is given in Tables \ref{tab:one}, \ref{tab:two}, and \ref{tab:three} respectively. 

\subsubsection{Gluonic excitations}
 \emph{Gluonic excitations} are hadrons with constituent gluonic fields. Since QCD-confined states are naturally nonperturbative, having constituent gluons inside hadrons is unsurprising. In this category, we have glueballs $G\,G$, hybrid mesons $q\,\bar{q}\, G$, and hybrid baryons $q\,q\,q\, G$, where $G$ is a constituent degree of flavor. In other words, a color-octet pair $q\, \bar{q}$ is color neutralized by an excited gluon $G$. These hybrid configurations allow for introducing other quantum numbers that are impossible in the quark constituent model, such as $J^{CP}=1^{+-}$ for hybrid vector mesons.

 Holographically, in the case of vector hybrid mesons, we will consider mesons with one gluon for the $Z_c$ candidates and two gluons for the $Z_b$ candidates. For a constituent gluon, $\Delta=5$ implies $M_5^2\, R^2=8$.For two gluons, $\Delta=7$ yields $M_5^2\, R^2=24$. The threshold masses for both cases are the following

 \begin{eqnarray*}
     \bar{m}_{Z_c}&=& 0.98\,m_c+0.2\,m_G=1.533\,\text{GeV} \\
     \bar{m}_{Z_b}&=& 0.99\,m_b+0.01\,m_G=4.6897\,\text{GeV}
 \end{eqnarray*}

 \noindent where we have assumed $m_q=m_{\bar{q}}$ implicitly. 
 
 A summary of the mass spectrum for these hybrid meson candidates can be found in Tables \ref{tab:one} for $Z_c$ and \ref{tab:three} for $Z_b$ states.
    
\begin{table*}
\centering
\begin{tabular}{c||c||c||c}
\hline
\multicolumn{4}{c}{\textbf{$Z_c$  states}}\\
\hline 
\hline
\textbf{Non-$q\bar{q}$ state} &\textbf{$Z_c(3900)$} &\textbf{$Z_c(4200)$)} &\textbf{$Z_c(4430)$}\\
\hline\hline
\textbf{Exp. Masses (MeV)}     & $3887.1\pm 3.6$ & $4296^{+35}_{-32}$ & $4478^{+13}_{-18}$   \\
\hline
\thead{\textbf{Diquark-Antidiquark (MeV)}\\ $\bar{m}_{4Q}=1550$ MeV  \\$\kappa =2151$ MeV and $\alpha=0.5387$} & $4004.8 \,(3.0\%)$ & $4384.9\,(2.1\%)$ & $4706.6\,(5.1\%)$   \\
\hline
\thead{\textbf{Hadronic Molecule (MeV)} \\ 
$\bar{m}_{HM}=1432.3$ MeV \\
$\kappa=1907$ MeV and $\alpha=0.4887$}& $3817.1\,(1.8\%)$& $4214.7\,(1.9\%)$ & $4552.1\,(1.6\%)$   \\
\hline
\thead{\textbf{Hybrid Meson (MeV)}  \\
$\bar{m}_{Z_c}=1533$ MeV\\
$\kappa=2114.8$ MeV and $\alpha=0.5317$}& $3721.9\,(4.2\%)$& $4156.4\,(3.2\%)$ &  $4513.2\,(0.8\%)$ \\
\hline
\hline
\end{tabular}
\caption{Holographic mass spectrum for the $Z_c$ non-$q\bar{q}$ states candidates according to their threshold mass $\bar{m}$. Parameters $\kappa$ and $\alpha$ are read from the expressions \eqref{kappa-1} and \eqref{alpha-1}. In the case of $Z_c$ as a hybrid meson, we consider one constituent gluon, i.e., $\Delta=5$. The quantity inside the parenthesis is the relative error.  Experimental data come from \cite{Workman:2022ynf}.}
\label{tab:one}
\end{table*}

\begin{table*}
\centering
\begin{tabular}{c||c||c||c}
\hline
\multicolumn{4}{c}{\textbf{$\psi$  states}}\\
\hline 
\hline
\textbf{Non-$q\bar{q}$ state} &\textbf{$\psi(4230)$} &\textbf{$\psi(4360)$)} &\textbf{$\psi(4660)$}\\
\hline\hline
\textbf{Exp. Masses (MeV)}     & $4222.5\pm 2.4$ & $4374 \pm 7$ & $4630\pm6$   \\
\hline
\thead{\textbf{Hadrocharmonium (MeV)}\\ $\bar{m}_{HQc}=1717$ MeV  \\$\kappa =2522.6$ MeV and $\alpha=0.6036$} & $4222.5 \,(0.2\%)$ & $4577.4\,(4.6\%)$ & $4871.8\,(5.2\%)$   \\
\hline
\thead{\textbf{Hadronic Molecule (MeV)} \\ 
$\bar{m}_{HM}=1549.1$ MeV \\
$\kappa=2149.2$ MeV and $\alpha=0.5384$}& $4003.5\,(5.2\%)$& $4383.7\,(0.2\%)$ & $4705.6\,(1.6\%)$   \\
\hline
\hline
\end{tabular}
\caption{Holographic mass spectrum for the $\psi$ (or $Y$) non-$q\bar{q}$ states candidates according to their threshold mass $\bar{m}$. Parameters $\kappa$ and $\alpha$ are read from the expressions \eqref{kappa-1} and \eqref{alpha-1}. The quantity inside the parenthesis is the relative error. The experimental data come from \cite{Workman:2022ynf}.}
\label{tab:two}
\end{table*}

\begin{table*}
\centering
\begin{tabular}{c||c||c}
\hline
\multicolumn{3}{c}{\textbf{$Z_b$  states}}\\
\hline 
\hline
\textbf{Non-$q\bar{q}$ state} &\textbf{$Z_b(10610)$} &\textbf{$Z_b(10650)$)} \\
\hline\hline
\textbf{Exp. Masses (MeV)}     & $10609\pm 6$ & $10652.2\pm 1.5$   \\
\hline
\thead{\textbf{Diquark-Antidiquark (MeV)}\\ $\bar{m}_{HM}=4753$ MeV  \\$\kappa =11269.5$ MeV and $\alpha=0.8633$} & $10224.5 \,(3.6\%)$ & $10517.6\,(1.3\%)$\\
\hline
\thead{\textbf{Hybrid Meson (MeV)} \\ 
$\bar{m}_{Zb}=4689.7$ MeV \\
$\kappa=11102.3$ MeV and $\alpha=0.8633$}& $10257.9\,(3.3\%)$& $10512.5\,(1.3\%)$\\
\hline
\hline
\end{tabular}
\caption{Holographic mass spectrum for the $Z_b$ non-$q\bar{q}$ states candidates according to their threshold mass $\bar{m}$. Parameters $\kappa$ and $\alpha$ are read from the expressions \eqref{kappa-1} and \eqref{alpha-1}. For $Z_b$ as a hybrid meson, we consider two constituent gluons, i.e., $\Delta=7$. The quantity inside the parenthesis is the relative error. Experimental data come from \cite{Workman:2022ynf}.}
\label{tab:three}
\end{table*}

\section{Differential Configurational Entropy}\label{entropy}

Configurational entropy (CE) constitutes a fundamental concept with far-reaching implications across various scientific disciplines, including statistical mechanics, information theory, and holography. This concept encapsulates the various arrangements or microstates that a given macrostate may embody. A higher CE value means a correspondingly greater multiplicity of potential microstate configurations. From a thermodynamic standpoint, CE refers to the work performed by a system with a nonexchange of temperature. Thus, CE could be helpful to infer when a given physical system, at zero temperature, is stable; we will say that the configuration with less CE is more stable than the others.  

In information theory, CE bridges the informational content of physical solutions to the equations of motion governing them. In particular, CE assumes the role of a \emph{logarithmic measure} quantifying the spatial complexity inherent in solutions localized within predefined energy boundaries. Conceptually, it quantifies the informational content inherent in solutions corresponding to a specified set of e.o.m. Consequently, CE may be construed as assessing the information required to characterize functions localized within a given parameter space comprehensively. In a broader context, dynamic solutions emerge from the extremization of actions, while CE functions as a metric capturing the underlying information content within these dynamically evolving states.

In the domain of discrete variables, CE is rigorously articulated as an application of Shannon entropy, succinctly formulated as \cite{Gleiser:2012tu, Bernardini:2016qit, Braga:2016wzx}:

\begin{equation}
    S_C=-\sum_n\,p_n\,\text{log}\,p_n.
\end{equation}

In the transition to continuous variables, the concept of differential configurational entropy (DCE) emerges, characterized by:

\begin{equation}\label{DCE-def}
S_C\left[f\right]=-\int{d^d\,k\,\tilde{f}\left(k\right)\,\text{log}\,\tilde{f}\left(k\right)},    
\end{equation}

\noindent where $\tilde{f}\left(k\right)=f\left(k\right)/f\left(k\right)_\text{Max}$. Here, $f(k)_\text{Max}$ represents the maximum value assumed by $f(k)$, and $f(k)$ itself belongs to the square-integrable space $L^2\left(\mathbb{R}^2\right)$, ensuring its Fourier transformability. Mathematically, $f(k)$ corresponds to the energy density within the momentum space, denoted as $\rho(k)$.

In the context of AdS/CFT duality, the application of CE finds fertile ground within both bottom-up and top-down AdS/QCD models, as initially delineated by \cite{Bernardini:2016hvx}. The versatility of CE becomes apparent in its applicability to a spectrum of physical scenarios, spanning from hadronic states to heavy quarkonium stability under varying thermal, magnetic, and density conditions. Furthermore, CE's utility extends to investigating the holographic deconfinement phase transition, as elucidated in \cite{Braga:2018fyc, Braga:2020hhs, Braga:2020myi, Braga:2020opg}, and recently in the exploration of holographic stability in light nuclides \cite{MartinContreras:2022lxl}.

Within this discourse, the holographic dictionary serves as a central construct, enabling the translation of spatial configurations of boundary particles into the holographic configuration of the dual bulk field. This process encapsulates the information encoded in the arrangement of constituents within a hadron, briefly represented by the energy density inherent in the temporal component of the energy-momentum tensor, denoted as $\rho(z) \equiv T_{00}(z)$.

The procedure for computing CE, as delineated by \cite{Bernardini:2016hvx, MartinContreras:2022lxl}, emanates from the  on-shell bulk action, yielding the energy-momentum tensor as

\begin{equation}\label{Energy-momentum}
T_{mn}=\frac{2}{\sqrt{-g}}\,\frac{\partial\left[\sqrt{-g}\,\mathcal{L}_\text{Hadron}\right]}{\partial\,g^{mn}}.
\end{equation}

From the holographic potential $V(z)$ \eqref{nonlinear-pot}
and inverting the transformation $\psi(z)=e^{B(z)/2}\,\phi(z)$, it is possible to obtain the Sturm-Liouville modes to feed up the on-shell energy-momentum tensor

\begin{multline}\label{energy-density}
 \rho(z)\equiv T_{00}=\frac{e^{-B\left(z\right)}}{2}\left(\frac{z}{R}\right)^3\times\\
\left\{ \left[\frac{1}{g_5^2}\left(M_n^2\,\psi_n^2+\psi_n'^2\right)-\frac{M_5^2\,R^2}{z^2}\psi_n^2\right] \right\}\,\Omega,
\end{multline}

The next step in the recipe for the CE is to define the \emph{modal fraction}. To do so is necessary to compute a Fourier-transformed representation of $\rho$, i.e., $\bar{\rho}(k)$. Recall that $\rho(z)\in L^2(\mathbb{R})$, and also has information on how energy is localized in the bulk.  Thus, it is an indirect measure of how normalizable modes are well localized in the AdS space. Therefore, from Plancherel's theorem, 

\begin{equation}
    \int_0^\infty{dz\,\left|\rho(z)\right|^2}=\int_0^\infty{dk\,\left|\bar{\rho}(k)\right|^2}, 
\end{equation}

\noindent it is possible to quantify \emph{localizabilty} by defining how spread the energy density is using the \emph{modal fraction}, defined as

\begin{equation}
   f(k) =\frac{ |\bar{\rho}(k)|^2}{\int dk |\bar{\rho}(k)|^2}.
\end{equation}

The differential configurational entropy (DCE) for the non-$q\,\bar{q}$ state is written from the modal fraction as

\begin{equation}
S_{DCE}=-\int dk \,\tilde{f}(k) \log \,\tilde{f}(k)
\end{equation}

\noindent where $\tilde{f}\left(k\right)=f\left(k\right)/f\left(k\right)_\text{Max}$. Notice that we normalize the modal fraction with $f\left(k\right)_\text{Max}$. 

In the following sections, we will use this tool to explore the stability of tetraquark candidates from their holographic duals constructed in the nonquadratic SWM. 

\section{Configurational entropy and hadron stability}  \label{hadron-stability}
Since the first applications of DCE to AdS/QCD, the connection between hadron stability and DCE has been open since there is no apparent direct connection from fundamental grounds. 

Stability in hadrons is an issue following confinement. Once constituents are bounded in a hadronic structure, it is natural to wonder if such a structure is stable enough (in time) to be considered a hadronic state.  

The connection with configurational entropy arises with the locality as a synonym for confinement: a QCD-bounded state of constituent quarks and gluons has highly well spatially localized wave functions for these constituents. Regarding the configurational entropy, these systems are more stable (less CE) than delocalized ones. In other words, confinement can be understood as a transition from localization to delocalization in space. 

Let us turn these ideas to the AdS/QCD realm. Mesons in bottom-up models appear from the field/operator duality: operators creating hadrons at the boundary are dual to bulk fields representing these hadrons. 

The connection is done with the operator dimension set as the bulk field scaling dimension (minus the spin), $\Delta$ \cite{Polchinski:2001tt, Balasubramanian:1998de}.  This information defines the bulk field mass $M_5$. Thus, for a given hadron, defined by an operator $\mathcal{O}$ with dimension $\Delta$ at the boundary, it will be dual to a bulk field with mass $M_5^2\, R^2=(\Delta-S)(\Delta+S-4)$. In other terms, \emph{a hadron in bottom-up models is a bag with $N$ constituents characterized by $M_5^2$}. Any information about the hadronic inner structure is not present \emph{ab initio}. 

This map between boundary and bulk physics also implies that normalizable bulk modes are dual to mesons at the boundary. From the AdS/CFT correspondence ground, this statement conflicts with field/operator duality since spatially localized wave functions at the boundary should be delocalized at the bulk because of the IR/UV behavior. However, this holds for $\mathcal{N}=4$ SYM theory dual to a Type IIB supergravity, where bulk states are unbounded \cite{Hamilton:2006az}.  Following \cite{Rehren:2000tp, Banks:1998dd}, one of the consequences of lifting the bulk conformal invariance (by placing a hard cutoff or using a dilaton field) is precisely this reinterpretation of the operator/field duality where extended and localized objects at the boundary are dual to localized objects at the bulk. This locality in the bulk objects seems natural in nonconformal theories \cite{Bena:1999jv}. This idea supports computing hadronic properties (mass spectrum, form factors, decay constants, thermal densities, etc.) in bottom-up models. 

Therefore, for bottom-up models (or models where nonconformality exists), local operators at the boundary can have local dual objects at the bulk. This affirmation connects with the configurational entropy directly. Thus, it is correct to infer that \emph{locality at the bulk (involving lower configurational entropy) implies stability at the boundary}. 

This hypothesis is strongly supported by the thermal or chemical analysis of these bottom-up models \cite{Braga:2016wzx, Braga:2017bml, MartinContreras:2021bis}, where the melting criterion is dictated by when the spectral peak disappears. This peak broadening can be understood as a thermal (chemical) delocalization of the bulk quasinormal modes. 

%At this point, then is appropriate to quote that normalizable bulk modes capture boundary meson physics in bottom-up models. A bulk normalizable mode, characterized by its bulk mass $M_5$, mimics hadrons as bags with constituents. If no inner configuration is provided, modes with the same bulk mass are equivalent despite the constituent content. One example is that a boundary scalar hadronic operator having $\Delta=6$ can be associated with pure gluon structures (4G) or hybrid ones with four quarks with a gluon. 
\begin{figure*}
  \includegraphics[width=2.4 in]{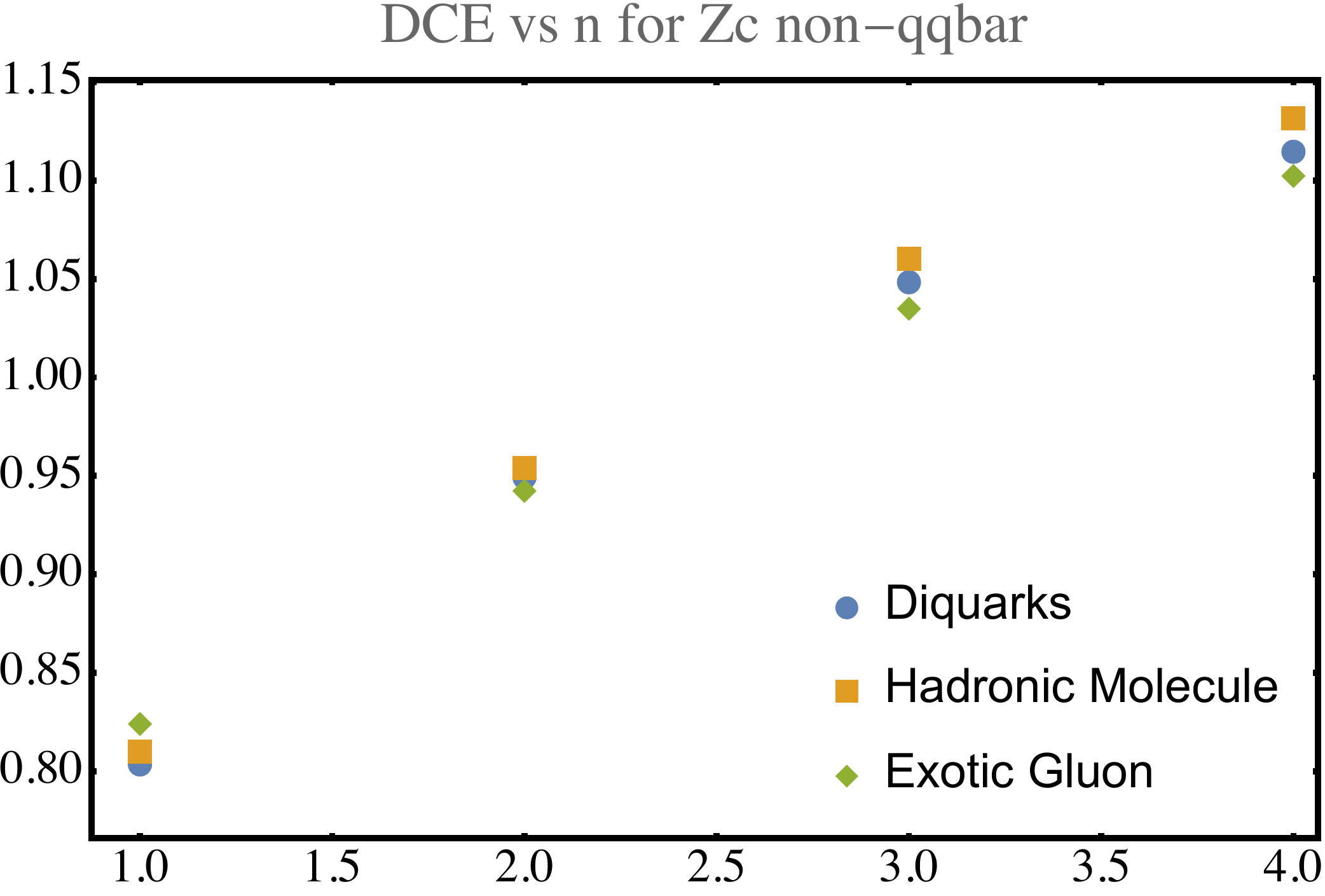}
  \includegraphics[width=2.4 in]{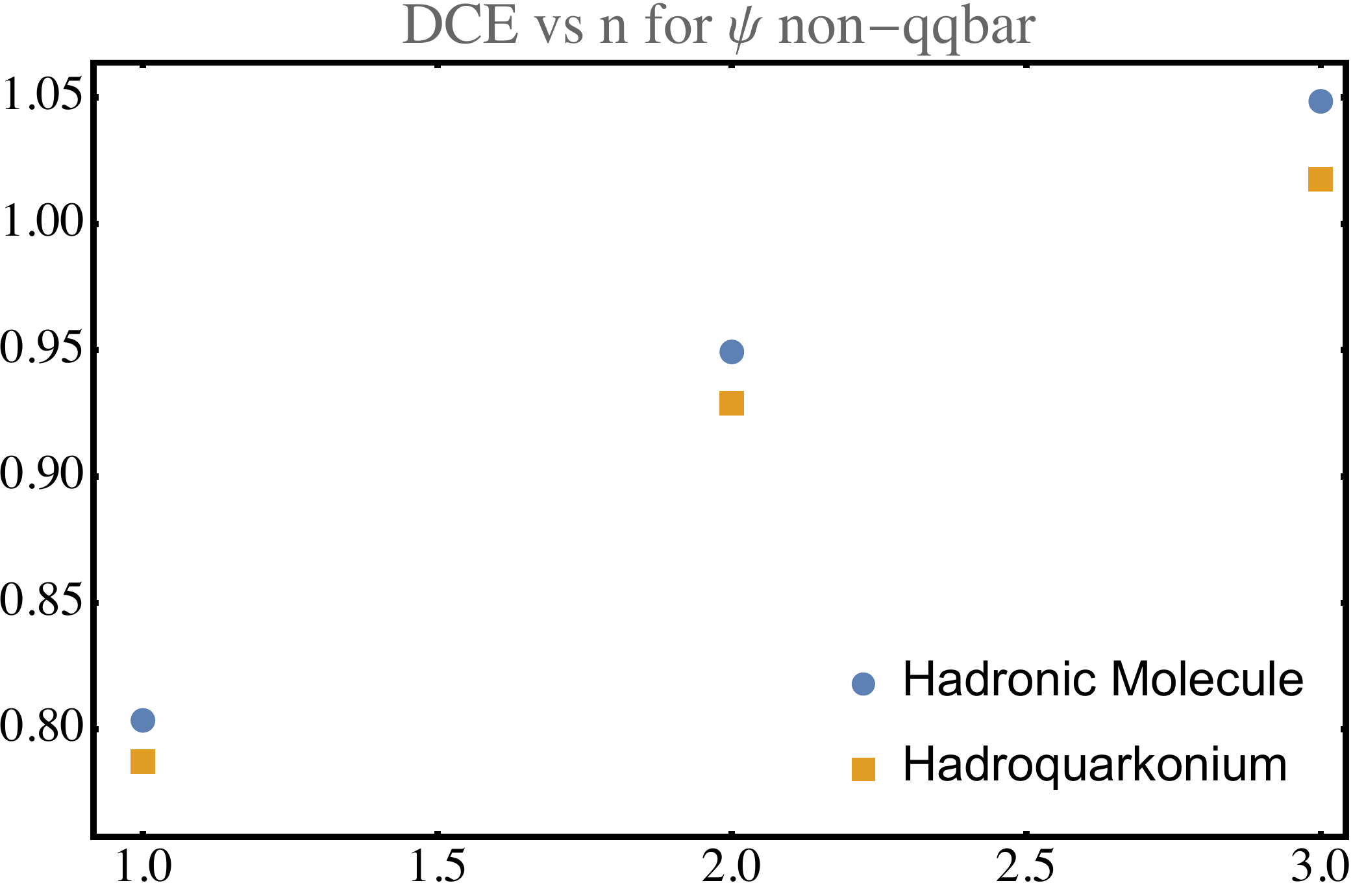}
  \includegraphics[width=2.4 in]{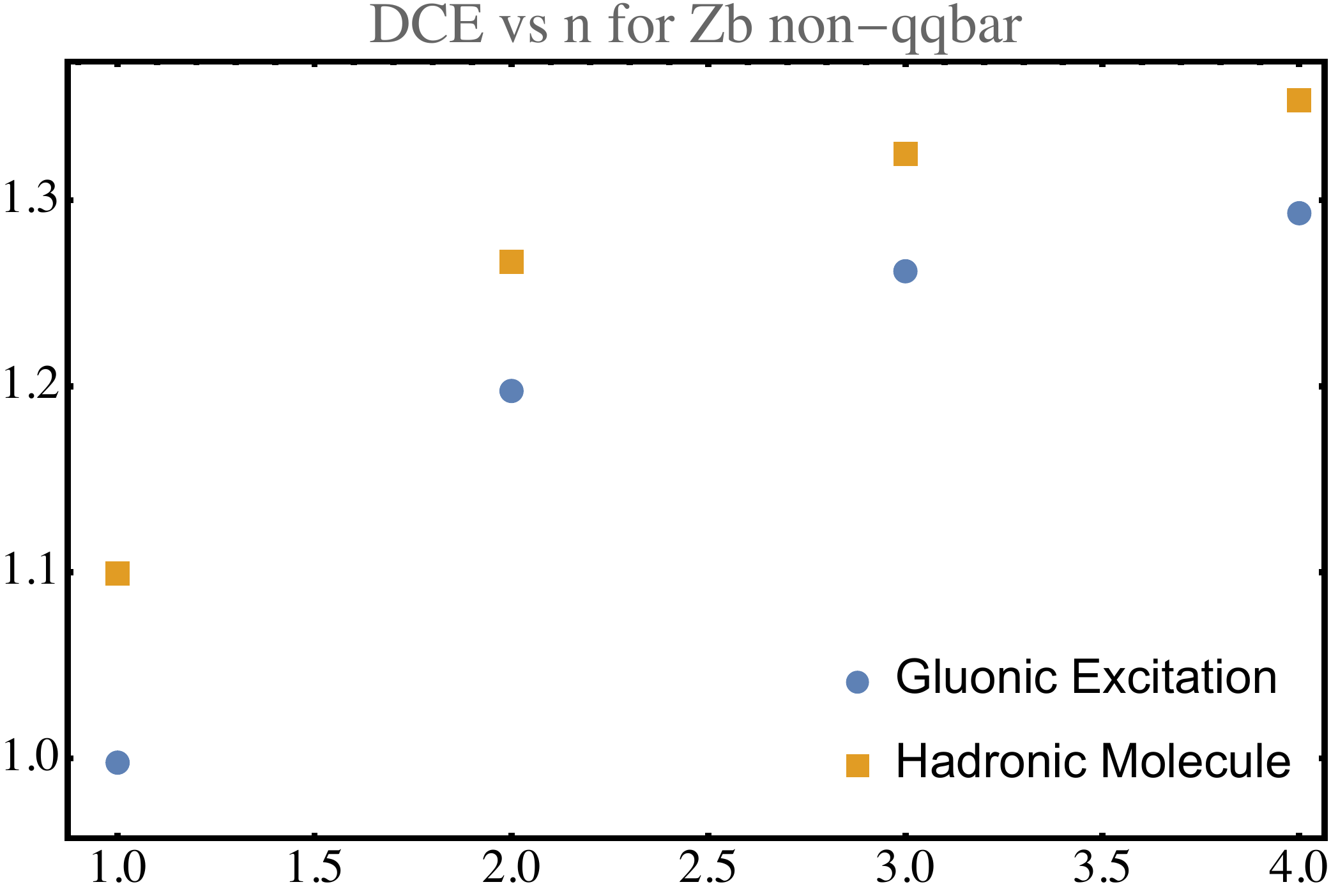}

\caption{Differential configurational entropy, in natural entropy units, as a function of excitation number for non-$q\,\bar{q}$ candidates. The left-upper panel plots the $Z_c$ holographic comparison between diquark, hadronic molecule, and exotic gluon structure. We present an analysis of $\psi$ or ($Y$) with hadronic molecule and hadrocharmonium structures in the right panel. The bottom panel presents the DCE for gluonic excitation (with two gluons) and hadronic molecule structures for $Z_b$ states. From DCE grounds, those structures with less entropy are the most stable and, therefore, preferred. }
\label{fig:one}
\end{figure*}

\section{DCE for non-$q\,\bar{q}$ states}\label{DCE-hadron}
We can discuss stability further once we have established that normalizable modes mimic hadrons at the boundary in locality terms. One of the main features of the nonquadratic dilaton is the inclusion of constituent mass dependence. Inspired by the well-known \cite{Chen:2018hnx} fact that the constituent mass affects the concavity of the Regge trajectory, that is, the linearity, this dilaton encloses the mass behavior in the quadratic deformation parameter $\alpha$. As in the original softwall proposal \cite{Karch:2006pv}, the slope $\kappa$ is still flavor dependent. However, this slope also runs with the constituent mass threshold $\bar{m}$. In practical terms, despite having two parameters at hand, we only have to deal with the mass threshold. Thus, at this point, we have increased the hadronic information encoded into the model since we have structural parameters included. 

In the original nonquadratic softwall work \cite{MartinContreras:2020cyg}, the holographic prediction of which non-$q\bar{q}$ structure was focused on minimizing the RMS error in the mass spectrum. Following this prescription, for $Z_c$, the lowest RMS error (1.5\%) is achieved in the hadronic molecule case. For $\psi$, the lowest error (2.7\%) is for the hadronic molecule case. And, for $Z_b$, the lowest RMS error (3.5\%) appears for hybrid meson (with two gluon flux tubes). Let us compare this criterion with the DCE analysis.

In the case of DCE, even though multiquark states for vector hadrons have the same bulk mass, the nonquadratic dilaton provides a mechanism to distinguish between structures through the threshold mass. This parametrization of the mass of the constituents allows the extrapolation of a pair $(\kappa,\alpha)$ for each structure. Tables \ref{tab:one}, \ref{tab:two}, and \ref{tab:three} summarize the extrapolated values for each structure, according to the non-$q\, \bar{q}$ candidate.  

The DCE analysis calculates the configurational entropy for each structure using the procedure exposed in Sec. \ref{entropy}. Following the idea that bulk modes capture locality information about hadron modes at the boundary, the trajectory with less DCE will be the most stable. This analysis differs entirely from the one done in \cite{Karapetyan:2021ufz}. In this work, the authors predicted the next generation of heavy non-$q\, \bar{q}$ by constructing Regge trajectories inspired by the DCE behavior of the associated bulk modes. They also discussed the possible existence of excited non-$q\,\bar{q}$ states from DCE holographic grounds, arguing that the lower production of these highly excited states agrees with the DCE holographic result, interpreted as hadronic information compressed, according to Shannon's information theory. Thus, only lower excitation modes are allowed and expected to dominate experiments. 

The analysis proposed in this work deals with the structural stability encoded in the threshold mass $\bar{m}$. Thus, we will compare the entropic effect of one parametrization over another for each family of states. The results of this analysis are summarized in Fig. \ref{fig:one}.

We compare three possible structures for the $Z_c$ states: diquark-antidiquark, hadronic molecule (multiquark states), and hybrid meson with one gluonic flux tube. For the ground state, the lowest DCE is achieved by the diquark structure. However, hybrid mesons (one exotic gluon flux tube) become more stable for excited states. Performing a quadrature analysis, i.e. looking for the smallest total DCE per structure, we found that the most stable configuration is the hybrid meson. 

We compare two multiquark structures in the $\psi$ states: hadronic molecule and hadrocharmonium. In this scenario, the smallest DCE is observed for hadrocharmonium.

Finally, for the $Z_b$ states, following hadron phenomenology\cite{Brambilla:2019esw}, we tested the hybrid meson (with two gluon tubes) and the hadronic molecule structure. We found the lowest DCE and, therefore, the most stable structure for the hadronic molecule.   

The DCE holographic analysis brings results different from the pure spectroscopic analysis, which does not account for the configurational information encoded into the threshold mass. 

\section{Conclusions}\label{conclusions}
As the vast and extensive literature demonstrates, in recent years, configurational entropy has proven to be an interesting tool to address in holographic models \cite{Colangelo:2018mrt, Karapetyan:2021ufz, Braga:2020myi, Braga:2020opg, Braga:2020hhs, Karapetyan:2021crv, Barreto:2022ohl, MartinContreras:2023eft}. This work explored the DCE to analyze stability when information regarding the hadronic inner structure is provided. In parallel, we also discussed the connection between hadronic stability and configurational entropy. Even thermodynamically speaking, CE does not account for processes where energy transfers change the configuration of a system. It can provide insights into configurational stability. Later, we can complement this discussion with thermal or chemical approaches to the concept.  

Speaking about hadronic stability is equivalent to talking about wave-function locality. This discussion stretches the hypothesis that normalizable modes living in the AdS bulk are dual to meson modes. This assumption is supported by slightly breaking the bulk conformal invariance, inducing bounded KK-towers dual to hadronic Regge trajectories. In bottom-up models, this is achieved by introducing a dilaton field. Thus, well-localized (bounded) bulk modes imply localized and confined objects at the boundary.

In this context, we found that, as a holographic prediction for heavy non-$q\,\bar{q}$ hadrons, the $Z_c$ states should be cataloged as \emph{hybrid meson}, $\psi$ states as \emph{hadrocharmonium}, and $Z_b$ as \emph{hadronic molecule}. 

\begin{acknowledgments}
M. A. Martin Contreras wants to acknowledge the financial support provided by the National Natural Science Foundation of China (NSFC) under Grant No 12350410371. 
\end{acknowledgments}

\bibliography{apssamp}% Produces the bibliography via BibTeX.

\providecommand{\noopsort}[1]{}\providecommand{\singleletter}[1]{#1}%
\begin{thebibliography}{70}%
\makeatletter
\providecommand \@ifxundefined [1]{%
 \@ifx{#1\undefined}
}%
\providecommand \@ifnum [1]{%
 \ifnum #1\expandafter \@firstoftwo
 \else \expandafter \@secondoftwo
 \fi
}%
\providecommand \@ifx [1]{%
 \ifx #1\expandafter \@firstoftwo
 \else \expandafter \@secondoftwo
 \fi
}%
\providecommand \natexlab [1]{#1}%
\providecommand \enquote  [1]{``#1''}%
\providecommand \bibnamefont  [1]{#1}%
\providecommand \bibfnamefont [1]{#1}%
\providecommand \citenamefont [1]{#1}%
\providecommand \href@noop [0]{\@secondoftwo}%
\providecommand \href [0]{\begingroup \@sanitize@url \@href}%
\providecommand \@href[1]{\@@startlink{#1}\@@href}%
\providecommand \@@href[1]{\endgroup#1\@@endlink}%
\providecommand \@sanitize@url [0]{\catcode `\\12\catcode `\$12\catcode
  `\&12\catcode `\#12\catcode `\^12\catcode `\_12\catcode `\%12\relax}%
\providecommand \@@startlink[1]{}%
\providecommand \@@endlink[0]{}%
\providecommand \url  [0]{\begingroup\@sanitize@url \@url }%
\providecommand \@url [1]{\endgroup\@href {#1}{\urlprefix }}%
\providecommand \urlprefix  [0]{URL }%
\providecommand \Eprint [0]{\href }%
\providecommand \doibase [0]{http://dx.doi.org/}%
\providecommand \selectlanguage [0]{\@gobble}%
\providecommand \bibinfo  [0]{\@secondoftwo}%
\providecommand \bibfield  [0]{\@secondoftwo}%
\providecommand \translation [1]{[#1]}%
\providecommand \BibitemOpen [0]{}%
\providecommand \bibitemStop [0]{}%
\providecommand \bibitemNoStop [0]{.\EOS\space}%
\providecommand \EOS [0]{\spacefactor3000\relax}%
\providecommand \BibitemShut  [1]{\csname bibitem#1\endcsname}%
\let\auto@bib@innerbib\@empty
%</preamble>
\bibitem [{\citenamefont {Karch}\ \emph {et~al.}(2006)\citenamefont {Karch},
  \citenamefont {Katz}, \citenamefont {Son},\ and\ \citenamefont
  {Stephanov}}]{Karch:2006pv}%
  \BibitemOpen
  \bibfield  {author} {\bibinfo {author} {\bibfnamefont {A.}~\bibnamefont
  {Karch}}, \bibinfo {author} {\bibfnamefont {E.}~\bibnamefont {Katz}},
  \bibinfo {author} {\bibfnamefont {D.~T.}\ \bibnamefont {Son}}, \ and\
  \bibinfo {author} {\bibfnamefont {M.~A.}\ \bibnamefont {Stephanov}},\ }\href
  {\doibase 10.1103/PhysRevD.74.015005} {\bibfield  {journal} {\bibinfo
  {journal} {Phys. Rev. D}\ }\textbf {\bibinfo {volume} {74}},\ \bibinfo
  {pages} {015005} (\bibinfo {year} {2006})},\ \Eprint
  {http://arxiv.org/abs/hep-ph/0602229} {arXiv:hep-ph/0602229} \BibitemShut
  {NoStop}%
\bibitem [{\citenamefont {Grigoryan}\ and\ \citenamefont
  {Radyushkin}(2007)}]{Grigoryan:2007my}%
  \BibitemOpen
  \bibfield  {author} {\bibinfo {author} {\bibfnamefont {H.~R.}\ \bibnamefont
  {Grigoryan}}\ and\ \bibinfo {author} {\bibfnamefont {A.~V.}\ \bibnamefont
  {Radyushkin}},\ }\href {\doibase 10.1103/PhysRevD.76.095007} {\bibfield
  {journal} {\bibinfo  {journal} {Phys. Rev. D}\ }\textbf {\bibinfo {volume}
  {76}},\ \bibinfo {pages} {095007} (\bibinfo {year} {2007})},\ \Eprint
  {http://arxiv.org/abs/0706.1543} {arXiv:0706.1543 [hep-ph]} \BibitemShut
  {NoStop}%
\bibitem [{\citenamefont {Abidin}\ and\ \citenamefont
  {Carlson}(2009)}]{Abidin:2009hr}%
  \BibitemOpen
  \bibfield  {author} {\bibinfo {author} {\bibfnamefont {Z.}~\bibnamefont
  {Abidin}}\ and\ \bibinfo {author} {\bibfnamefont {C.~E.}\ \bibnamefont
  {Carlson}},\ }\href {\doibase 10.1103/PhysRevD.79.115003} {\bibfield
  {journal} {\bibinfo  {journal} {Phys. Rev. D}\ }\textbf {\bibinfo {volume}
  {79}},\ \bibinfo {pages} {115003} (\bibinfo {year} {2009})},\ \Eprint
  {http://arxiv.org/abs/0903.4818} {arXiv:0903.4818 [hep-ph]} \BibitemShut
  {NoStop}%
\bibitem [{\citenamefont {Herzog}(2007)}]{Herzog:2006ra}%
  \BibitemOpen
  \bibfield  {author} {\bibinfo {author} {\bibfnamefont {C.~P.}\ \bibnamefont
  {Herzog}},\ }\href {\doibase 10.1103/PhysRevLett.98.091601} {\bibfield
  {journal} {\bibinfo  {journal} {Phys. Rev. Lett.}\ }\textbf {\bibinfo
  {volume} {98}},\ \bibinfo {pages} {091601} (\bibinfo {year} {2007})},\
  \Eprint {http://arxiv.org/abs/hep-th/0608151} {arXiv:hep-th/0608151}
  \BibitemShut {NoStop}%
\bibitem [{\citenamefont {Cai}\ and\ \citenamefont {Shock}(2007)}]{Cai:2007zw}%
  \BibitemOpen
  \bibfield  {author} {\bibinfo {author} {\bibfnamefont {R.-G.}\ \bibnamefont
  {Cai}}\ and\ \bibinfo {author} {\bibfnamefont {J.~P.}\ \bibnamefont
  {Shock}},\ }\href {\doibase 10.1088/1126-6708/2007/08/095} {\bibfield
  {journal} {\bibinfo  {journal} {JHEP}\ }\textbf {\bibinfo {volume} {08}},\
  \bibinfo {pages} {095} (\bibinfo {year} {2007})},\ \Eprint
  {http://arxiv.org/abs/0705.3388} {arXiv:0705.3388 [hep-th]} \BibitemShut
  {NoStop}%
\bibitem [{\citenamefont {Afonin}\ and\ \citenamefont
  {Katanaeva}(2018)}]{Afonin:2018era}%
  \BibitemOpen
  \bibfield  {author} {\bibinfo {author} {\bibfnamefont {S.~S.}\ \bibnamefont
  {Afonin}}\ and\ \bibinfo {author} {\bibfnamefont {A.~D.}\ \bibnamefont
  {Katanaeva}},\ }\href {\doibase 10.1103/PhysRevD.98.114027} {\bibfield
  {journal} {\bibinfo  {journal} {Phys. Rev. D}\ }\textbf {\bibinfo {volume}
  {98}},\ \bibinfo {pages} {114027} (\bibinfo {year} {2018})},\ \Eprint
  {http://arxiv.org/abs/1809.07730} {arXiv:1809.07730 [hep-ph]} \BibitemShut
  {NoStop}%
\bibitem [{\citenamefont {Ballon-Bayona}\ \emph {et~al.}(2021)\citenamefont
  {Ballon-Bayona}, \citenamefont {Mamani},\ and\ \citenamefont
  {Rodrigues}}]{Ballon-Bayona:2021ibm}%
  \BibitemOpen
  \bibfield  {author} {\bibinfo {author} {\bibfnamefont {A.}~\bibnamefont
  {Ballon-Bayona}}, \bibinfo {author} {\bibfnamefont {L.~A.~H.}\ \bibnamefont
  {Mamani}}, \ and\ \bibinfo {author} {\bibfnamefont {D.~M.}\ \bibnamefont
  {Rodrigues}},\ }\href {\doibase 10.1103/PhysRevD.104.126029} {\bibfield
  {journal} {\bibinfo  {journal} {Phys. Rev. D}\ }\textbf {\bibinfo {volume}
  {104}},\ \bibinfo {pages} {126029} (\bibinfo {year} {2021})},\ \Eprint
  {http://arxiv.org/abs/2107.10983} {arXiv:2107.10983 [hep-ph]} \BibitemShut
  {NoStop}%
\bibitem [{\citenamefont {Ballon~Bayona}\ \emph {et~al.}(2008)\citenamefont
  {Ballon~Bayona}, \citenamefont {Boschi-Filho},\ and\ \citenamefont
  {Braga}}]{BallonBayona:2007qr}%
  \BibitemOpen
  \bibfield  {author} {\bibinfo {author} {\bibfnamefont {C.~A.}\ \bibnamefont
  {Ballon~Bayona}}, \bibinfo {author} {\bibfnamefont {H.}~\bibnamefont
  {Boschi-Filho}}, \ and\ \bibinfo {author} {\bibfnamefont {N.~R.~F.}\
  \bibnamefont {Braga}},\ }\href {\doibase 10.1088/1126-6708/2008/03/064}
  {\bibfield  {journal} {\bibinfo  {journal} {JHEP}\ }\textbf {\bibinfo
  {volume} {03}},\ \bibinfo {pages} {064} (\bibinfo {year} {2008})},\ \Eprint
  {http://arxiv.org/abs/0711.0221} {arXiv:0711.0221 [hep-th]} \BibitemShut
  {NoStop}%
\bibitem [{\citenamefont {Braga}\ and\ \citenamefont
  {Vega}(2012)}]{Braga:2011wa}%
  \BibitemOpen
  \bibfield  {author} {\bibinfo {author} {\bibfnamefont {N.~R.~F.}\
  \bibnamefont {Braga}}\ and\ \bibinfo {author} {\bibfnamefont
  {A.}~\bibnamefont {Vega}},\ }\href {\doibase 10.1140/epjc/s10052-012-2236-2}
  {\bibfield  {journal} {\bibinfo  {journal} {Eur. Phys. J. C}\ }\textbf
  {\bibinfo {volume} {72}},\ \bibinfo {pages} {2236} (\bibinfo {year}
  {2012})},\ \Eprint {http://arxiv.org/abs/1110.2548} {arXiv:1110.2548
  [hep-ph]} \BibitemShut {NoStop}%
\bibitem [{\citenamefont {Vega}\ \emph {et~al.}(2011)\citenamefont {Vega},
  \citenamefont {Schmidt}, \citenamefont {Gutsche},\ and\ \citenamefont
  {Lyubovitskij}}]{Vega:2010ns}%
  \BibitemOpen
  \bibfield  {author} {\bibinfo {author} {\bibfnamefont {A.}~\bibnamefont
  {Vega}}, \bibinfo {author} {\bibfnamefont {I.}~\bibnamefont {Schmidt}},
  \bibinfo {author} {\bibfnamefont {T.}~\bibnamefont {Gutsche}}, \ and\
  \bibinfo {author} {\bibfnamefont {V.~E.}\ \bibnamefont {Lyubovitskij}},\
  }\href {\doibase 10.1103/PhysRevD.83.036001} {\bibfield  {journal} {\bibinfo
  {journal} {Phys. Rev. D}\ }\textbf {\bibinfo {volume} {83}},\ \bibinfo
  {pages} {036001} (\bibinfo {year} {2011})},\ \Eprint
  {http://arxiv.org/abs/1010.2815} {arXiv:1010.2815 [hep-ph]} \BibitemShut
  {NoStop}%
\bibitem [{\citenamefont {Colangelo}\ \emph {et~al.}(2008)\citenamefont
  {Colangelo}, \citenamefont {De~Fazio}, \citenamefont {Giannuzzi},
  \citenamefont {Jugeau},\ and\ \citenamefont {Nicotri}}]{Colangelo:2008us}%
  \BibitemOpen
  \bibfield  {author} {\bibinfo {author} {\bibfnamefont {P.}~\bibnamefont
  {Colangelo}}, \bibinfo {author} {\bibfnamefont {F.}~\bibnamefont {De~Fazio}},
  \bibinfo {author} {\bibfnamefont {F.}~\bibnamefont {Giannuzzi}}, \bibinfo
  {author} {\bibfnamefont {F.}~\bibnamefont {Jugeau}}, \ and\ \bibinfo {author}
  {\bibfnamefont {S.}~\bibnamefont {Nicotri}},\ }\href {\doibase
  10.1103/PhysRevD.78.055009} {\bibfield  {journal} {\bibinfo  {journal} {Phys.
  Rev. D}\ }\textbf {\bibinfo {volume} {78}},\ \bibinfo {pages} {055009}
  (\bibinfo {year} {2008})},\ \Eprint {http://arxiv.org/abs/0807.1054}
  {arXiv:0807.1054 [hep-ph]} \BibitemShut {NoStop}%
\bibitem [{\citenamefont {Vega}\ and\ \citenamefont
  {Schmidt}(2008)}]{Vega:2008af}%
  \BibitemOpen
  \bibfield  {author} {\bibinfo {author} {\bibfnamefont {A.}~\bibnamefont
  {Vega}}\ and\ \bibinfo {author} {\bibfnamefont {I.}~\bibnamefont {Schmidt}},\
  }\href {\doibase 10.1103/PhysRevD.78.017703} {\bibfield  {journal} {\bibinfo
  {journal} {Phys. Rev. D}\ }\textbf {\bibinfo {volume} {78}},\ \bibinfo
  {pages} {017703} (\bibinfo {year} {2008})},\ \Eprint
  {http://arxiv.org/abs/0806.2267} {arXiv:0806.2267 [hep-ph]} \BibitemShut
  {NoStop}%
\bibitem [{\citenamefont {Vega}\ and\ \citenamefont
  {Schmidt}(2009)}]{Vega:2008te}%
  \BibitemOpen
  \bibfield  {author} {\bibinfo {author} {\bibfnamefont {A.}~\bibnamefont
  {Vega}}\ and\ \bibinfo {author} {\bibfnamefont {I.}~\bibnamefont {Schmidt}},\
  }\href {\doibase 10.1103/PhysRevD.79.055003} {\bibfield  {journal} {\bibinfo
  {journal} {Phys. Rev. D}\ }\textbf {\bibinfo {volume} {79}},\ \bibinfo
  {pages} {055003} (\bibinfo {year} {2009})},\ \Eprint
  {http://arxiv.org/abs/0811.4638} {arXiv:0811.4638 [hep-ph]} \BibitemShut
  {NoStop}%
\bibitem [{\citenamefont {Branz}\ \emph {et~al.}(2010)\citenamefont {Branz},
  \citenamefont {Gutsche}, \citenamefont {Lyubovitskij}, \citenamefont
  {Schmidt},\ and\ \citenamefont {Vega}}]{Branz:2010ub}%
  \BibitemOpen
  \bibfield  {author} {\bibinfo {author} {\bibfnamefont {T.}~\bibnamefont
  {Branz}}, \bibinfo {author} {\bibfnamefont {T.}~\bibnamefont {Gutsche}},
  \bibinfo {author} {\bibfnamefont {V.~E.}\ \bibnamefont {Lyubovitskij}},
  \bibinfo {author} {\bibfnamefont {I.}~\bibnamefont {Schmidt}}, \ and\
  \bibinfo {author} {\bibfnamefont {A.}~\bibnamefont {Vega}},\ }\href {\doibase
  10.1103/PhysRevD.82.074022} {\bibfield  {journal} {\bibinfo  {journal} {Phys.
  Rev. D}\ }\textbf {\bibinfo {volume} {82}},\ \bibinfo {pages} {074022}
  (\bibinfo {year} {2010})},\ \Eprint {http://arxiv.org/abs/1008.0268}
  {arXiv:1008.0268 [hep-ph]} \BibitemShut {NoStop}%
\bibitem [{\citenamefont {Gutsche}\ \emph {et~al.}(2013)\citenamefont
  {Gutsche}, \citenamefont {Lyubovitskij}, \citenamefont {Schmidt},\ and\
  \citenamefont {Vega}}]{Gutsche:2012wb}%
  \BibitemOpen
  \bibfield  {author} {\bibinfo {author} {\bibfnamefont {T.}~\bibnamefont
  {Gutsche}}, \bibinfo {author} {\bibfnamefont {V.~E.}\ \bibnamefont
  {Lyubovitskij}}, \bibinfo {author} {\bibfnamefont {I.}~\bibnamefont
  {Schmidt}}, \ and\ \bibinfo {author} {\bibfnamefont {A.}~\bibnamefont
  {Vega}},\ }\href {\doibase 10.1103/PhysRevD.87.016017} {\bibfield  {journal}
  {\bibinfo  {journal} {Phys. Rev. D}\ }\textbf {\bibinfo {volume} {87}},\
  \bibinfo {pages} {016017} (\bibinfo {year} {2013})},\ \Eprint
  {http://arxiv.org/abs/1212.6252} {arXiv:1212.6252 [hep-ph]} \BibitemShut
  {NoStop}%
\bibitem [{\citenamefont {Batell}\ and\ \citenamefont
  {Gherghetta}(2008)}]{Batell:2008zm}%
  \BibitemOpen
  \bibfield  {author} {\bibinfo {author} {\bibfnamefont {B.}~\bibnamefont
  {Batell}}\ and\ \bibinfo {author} {\bibfnamefont {T.}~\bibnamefont
  {Gherghetta}},\ }\href {\doibase 10.1103/PhysRevD.78.026002} {\bibfield
  {journal} {\bibinfo  {journal} {Phys. Rev. D}\ }\textbf {\bibinfo {volume}
  {78}},\ \bibinfo {pages} {026002} (\bibinfo {year} {2008})},\ \Eprint
  {http://arxiv.org/abs/0801.4383} {arXiv:0801.4383 [hep-ph]} \BibitemShut
  {NoStop}%
\bibitem [{\citenamefont {Li}\ \emph {et~al.}(2013)\citenamefont {Li},
  \citenamefont {Huang},\ and\ \citenamefont {Yan}}]{Li:2012ay}%
  \BibitemOpen
  \bibfield  {author} {\bibinfo {author} {\bibfnamefont {D.}~\bibnamefont
  {Li}}, \bibinfo {author} {\bibfnamefont {M.}~\bibnamefont {Huang}}, \ and\
  \bibinfo {author} {\bibfnamefont {Q.-S.}\ \bibnamefont {Yan}},\ }\href
  {\doibase 10.1140/epjc/s10052-013-2615-3} {\bibfield  {journal} {\bibinfo
  {journal} {Eur. Phys. J. C}\ }\textbf {\bibinfo {volume} {73}},\ \bibinfo
  {pages} {2615} (\bibinfo {year} {2013})},\ \Eprint
  {http://arxiv.org/abs/1206.2824} {arXiv:1206.2824 [hep-th]} \BibitemShut
  {NoStop}%
\bibitem [{\citenamefont {Li}\ and\ \citenamefont {Huang}(2013)}]{Li:2013oda}%
  \BibitemOpen
  \bibfield  {author} {\bibinfo {author} {\bibfnamefont {D.}~\bibnamefont
  {Li}}\ and\ \bibinfo {author} {\bibfnamefont {M.}~\bibnamefont {Huang}},\
  }\href {\doibase 10.1007/JHEP11(2013)088} {\bibfield  {journal} {\bibinfo
  {journal} {JHEP}\ }\textbf {\bibinfo {volume} {11}},\ \bibinfo {pages} {088}
  (\bibinfo {year} {2013})},\ \Eprint {http://arxiv.org/abs/1303.6929}
  {arXiv:1303.6929 [hep-ph]} \BibitemShut {NoStop}%
\bibitem [{\citenamefont {Colangelo}\ \emph {et~al.}(2011)\citenamefont
  {Colangelo}, \citenamefont {Giannuzzi},\ and\ \citenamefont
  {Nicotri}}]{Colangelo:2010pe}%
  \BibitemOpen
  \bibfield  {author} {\bibinfo {author} {\bibfnamefont {P.}~\bibnamefont
  {Colangelo}}, \bibinfo {author} {\bibfnamefont {F.}~\bibnamefont
  {Giannuzzi}}, \ and\ \bibinfo {author} {\bibfnamefont {S.}~\bibnamefont
  {Nicotri}},\ }\href {\doibase 10.1103/PhysRevD.83.035015} {\bibfield
  {journal} {\bibinfo  {journal} {Phys. Rev. D}\ }\textbf {\bibinfo {volume}
  {83}},\ \bibinfo {pages} {035015} (\bibinfo {year} {2011})},\ \Eprint
  {http://arxiv.org/abs/1008.3116} {arXiv:1008.3116 [hep-ph]} \BibitemShut
  {NoStop}%
\bibitem [{\citenamefont {Sachan}\ and\ \citenamefont
  {Siwach}(2012)}]{Sachan:2011iy}%
  \BibitemOpen
  \bibfield  {author} {\bibinfo {author} {\bibfnamefont {S.}~\bibnamefont
  {Sachan}}\ and\ \bibinfo {author} {\bibfnamefont {S.}~\bibnamefont
  {Siwach}},\ }\href {\doibase 10.1142/S0217732312501635} {\bibfield  {journal}
  {\bibinfo  {journal} {Mod. Phys. Lett. A}\ }\textbf {\bibinfo {volume}
  {27}},\ \bibinfo {pages} {1250163} (\bibinfo {year} {2012})},\ \Eprint
  {http://arxiv.org/abs/1109.5523} {arXiv:1109.5523 [hep-th]} \BibitemShut
  {NoStop}%
\bibitem [{\citenamefont {Kim}\ \emph {et~al.}(2007)\citenamefont {Kim},
  \citenamefont {Lee},\ and\ \citenamefont {Lee}}]{Kim:2007rt}%
  \BibitemOpen
  \bibfield  {author} {\bibinfo {author} {\bibfnamefont {Y.}~\bibnamefont
  {Kim}}, \bibinfo {author} {\bibfnamefont {J.-P.}\ \bibnamefont {Lee}}, \ and\
  \bibinfo {author} {\bibfnamefont {S.~H.}\ \bibnamefont {Lee}},\ }\href
  {\doibase 10.1103/PhysRevD.75.114008} {\bibfield  {journal} {\bibinfo
  {journal} {Phys. Rev. D}\ }\textbf {\bibinfo {volume} {75}},\ \bibinfo
  {pages} {114008} (\bibinfo {year} {2007})},\ \Eprint
  {http://arxiv.org/abs/hep-ph/0703172} {arXiv:hep-ph/0703172} \BibitemShut
  {NoStop}%
\bibitem [{\citenamefont {Fujita}\ \emph {et~al.}(2010)\citenamefont {Fujita},
  \citenamefont {Kikuchi}, \citenamefont {Fukushima}, \citenamefont {Misumi},\
  and\ \citenamefont {Murata}}]{Fujita:2009ca}%
  \BibitemOpen
  \bibfield  {author} {\bibinfo {author} {\bibfnamefont {M.}~\bibnamefont
  {Fujita}}, \bibinfo {author} {\bibfnamefont {T.}~\bibnamefont {Kikuchi}},
  \bibinfo {author} {\bibfnamefont {K.}~\bibnamefont {Fukushima}}, \bibinfo
  {author} {\bibfnamefont {T.}~\bibnamefont {Misumi}}, \ and\ \bibinfo {author}
  {\bibfnamefont {M.}~\bibnamefont {Murata}},\ }\href {\doibase
  10.1103/PhysRevD.81.065024} {\bibfield  {journal} {\bibinfo  {journal} {Phys.
  Rev. D}\ }\textbf {\bibinfo {volume} {81}},\ \bibinfo {pages} {065024}
  (\bibinfo {year} {2010})},\ \Eprint {http://arxiv.org/abs/0911.2298}
  {arXiv:0911.2298 [hep-ph]} \BibitemShut {NoStop}%
\bibitem [{\citenamefont {Fujita}\ \emph {et~al.}(2009)\citenamefont {Fujita},
  \citenamefont {Fukushima}, \citenamefont {Misumi},\ and\ \citenamefont
  {Murata}}]{Fujita:2009wc}%
  \BibitemOpen
  \bibfield  {author} {\bibinfo {author} {\bibfnamefont {M.}~\bibnamefont
  {Fujita}}, \bibinfo {author} {\bibfnamefont {K.}~\bibnamefont {Fukushima}},
  \bibinfo {author} {\bibfnamefont {T.}~\bibnamefont {Misumi}}, \ and\ \bibinfo
  {author} {\bibfnamefont {M.}~\bibnamefont {Murata}},\ }\href {\doibase
  10.1103/PhysRevD.80.035001} {\bibfield  {journal} {\bibinfo  {journal} {Phys.
  Rev. D}\ }\textbf {\bibinfo {volume} {80}},\ \bibinfo {pages} {035001}
  (\bibinfo {year} {2009})},\ \Eprint {http://arxiv.org/abs/0903.2316}
  {arXiv:0903.2316 [hep-ph]} \BibitemShut {NoStop}%
\bibitem [{\citenamefont {Dudal}\ \emph {et~al.}(2016)\citenamefont {Dudal},
  \citenamefont {Granado},\ and\ \citenamefont {Mertens}}]{Dudal:2015wfn}%
  \BibitemOpen
  \bibfield  {author} {\bibinfo {author} {\bibfnamefont {D.}~\bibnamefont
  {Dudal}}, \bibinfo {author} {\bibfnamefont {D.~R.}\ \bibnamefont {Granado}},
  \ and\ \bibinfo {author} {\bibfnamefont {T.~G.}\ \bibnamefont {Mertens}},\
  }\href {\doibase 10.1103/PhysRevD.93.125004} {\bibfield  {journal} {\bibinfo
  {journal} {Phys. Rev. D}\ }\textbf {\bibinfo {volume} {93}},\ \bibinfo
  {pages} {125004} (\bibinfo {year} {2016})},\ \Eprint
  {http://arxiv.org/abs/1511.04042} {arXiv:1511.04042 [hep-th]} \BibitemShut
  {NoStop}%
\bibitem [{\citenamefont {Li}\ \emph {et~al.}(2017)\citenamefont {Li},
  \citenamefont {Huang}, \citenamefont {Yang},\ and\ \citenamefont
  {Yuan}}]{Li:2016gfn}%
  \BibitemOpen
  \bibfield  {author} {\bibinfo {author} {\bibfnamefont {D.}~\bibnamefont
  {Li}}, \bibinfo {author} {\bibfnamefont {M.}~\bibnamefont {Huang}}, \bibinfo
  {author} {\bibfnamefont {Y.}~\bibnamefont {Yang}}, \ and\ \bibinfo {author}
  {\bibfnamefont {P.-H.}\ \bibnamefont {Yuan}},\ }\href {\doibase
  10.1007/JHEP02(2017)030} {\bibfield  {journal} {\bibinfo  {journal} {JHEP}\
  }\textbf {\bibinfo {volume} {02}},\ \bibinfo {pages} {030} (\bibinfo {year}
  {2017})},\ \Eprint {http://arxiv.org/abs/1610.04618} {arXiv:1610.04618
  [hep-th]} \BibitemShut {NoStop}%
\bibitem [{\citenamefont {Fang}\ \emph {et~al.}(2021)\citenamefont {Fang},
  \citenamefont {Li},\ and\ \citenamefont {Wu}}]{Fang:2021ucy}%
  \BibitemOpen
  \bibfield  {author} {\bibinfo {author} {\bibfnamefont {Z.}~\bibnamefont
  {Fang}}, \bibinfo {author} {\bibfnamefont {Y.-Y.}\ \bibnamefont {Li}}, \ and\
  \bibinfo {author} {\bibfnamefont {Y.-L.}\ \bibnamefont {Wu}},\ }\href
  {\doibase 10.1140/epjc/s10052-021-09311-5} {\bibfield  {journal} {\bibinfo
  {journal} {Eur. Phys. J. C}\ }\textbf {\bibinfo {volume} {81}},\ \bibinfo
  {pages} {545} (\bibinfo {year} {2021})}\BibitemShut {NoStop}%
\bibitem [{\citenamefont {Chen}(2018{\natexlab{a}})}]{Chen:2018nnr}%
  \BibitemOpen
  \bibfield  {author} {\bibinfo {author} {\bibfnamefont {J.-K.}\ \bibnamefont
  {Chen}},\ }\href {\doibase 10.1140/epjc/s10052-018-6134-0} {\bibfield
  {journal} {\bibinfo  {journal} {Eur. Phys. J. C}\ }\textbf {\bibinfo {volume}
  {78}},\ \bibinfo {pages} {648} (\bibinfo {year}
  {2018}{\natexlab{a}})}\BibitemShut {NoStop}%
\bibitem [{\citenamefont {Martin~Contreras}\ and\ \citenamefont
  {Vega}(2020)}]{MartinContreras:2020cyg}%
  \BibitemOpen
  \bibfield  {author} {\bibinfo {author} {\bibfnamefont {M.~A.}\ \bibnamefont
  {Martin~Contreras}}\ and\ \bibinfo {author} {\bibfnamefont {A.}~\bibnamefont
  {Vega}},\ }\href {\doibase 10.1103/PhysRevD.102.046007} {\bibfield  {journal}
  {\bibinfo  {journal} {Phys. Rev. D}\ }\textbf {\bibinfo {volume} {102}},\
  \bibinfo {pages} {046007} (\bibinfo {year} {2020})},\ \Eprint
  {http://arxiv.org/abs/2004.10286} {arXiv:2004.10286 [hep-ph]} \BibitemShut
  {NoStop}%
\bibitem [{\citenamefont {Colangelo}\ and\ \citenamefont
  {Loparco}(2019)}]{Colangelo:2018mrt}%
  \BibitemOpen
  \bibfield  {author} {\bibinfo {author} {\bibfnamefont {P.}~\bibnamefont
  {Colangelo}}\ and\ \bibinfo {author} {\bibfnamefont {F.}~\bibnamefont
  {Loparco}},\ }\href {\doibase 10.1016/j.physletb.2018.11.053} {\bibfield
  {journal} {\bibinfo  {journal} {Phys. Lett. B}\ }\textbf {\bibinfo {volume}
  {788}},\ \bibinfo {pages} {500} (\bibinfo {year} {2019})},\ \Eprint
  {http://arxiv.org/abs/1811.05272} {arXiv:1811.05272 [hep-ph]} \BibitemShut
  {NoStop}%
\bibitem [{\citenamefont {Karapetyan}\ and\ \citenamefont
  {da~Rocha}(2021)}]{Karapetyan:2021ufz}%
  \BibitemOpen
  \bibfield  {author} {\bibinfo {author} {\bibfnamefont {G.}~\bibnamefont
  {Karapetyan}}\ and\ \bibinfo {author} {\bibfnamefont {R.}~\bibnamefont
  {da~Rocha}},\ }\href {\doibase 10.1140/epjp/s13360-021-01942-7} {\bibfield
  {journal} {\bibinfo  {journal} {Eur. Phys. J. Plus}\ }\textbf {\bibinfo
  {volume} {136}},\ \bibinfo {pages} {993} (\bibinfo {year} {2021})},\ \Eprint
  {http://arxiv.org/abs/2103.10863} {arXiv:2103.10863 [hep-ph]} \BibitemShut
  {NoStop}%
\bibitem [{\citenamefont {Chen}(2018{\natexlab{b}})}]{Chen:2018bbr}%
  \BibitemOpen
  \bibfield  {author} {\bibinfo {author} {\bibfnamefont {J.-K.}\ \bibnamefont
  {Chen}},\ }\href {\doibase 10.1016/j.physletb.2018.10.022} {\bibfield
  {journal} {\bibinfo  {journal} {Phys. Lett. B}\ }\textbf {\bibinfo {volume}
  {786}},\ \bibinfo {pages} {477} (\bibinfo {year} {2018}{\natexlab{b}})},\
  \Eprint {http://arxiv.org/abs/1807.11003} {arXiv:1807.11003 [hep-ph]}
  \BibitemShut {NoStop}%
\bibitem [{\citenamefont {Afonin}\ and\ \citenamefont
  {Pusenkov}(2014)}]{Afonin:2014nya}%
  \BibitemOpen
  \bibfield  {author} {\bibinfo {author} {\bibfnamefont {S.~S.}\ \bibnamefont
  {Afonin}}\ and\ \bibinfo {author} {\bibfnamefont {I.~V.}\ \bibnamefont
  {Pusenkov}},\ }\href {\doibase 10.1103/PhysRevD.90.094020} {\bibfield
  {journal} {\bibinfo  {journal} {Phys. Rev. D}\ }\textbf {\bibinfo {volume}
  {90}},\ \bibinfo {pages} {094020} (\bibinfo {year} {2014})},\ \Eprint
  {http://arxiv.org/abs/1411.2390} {arXiv:1411.2390 [hep-ph]} \BibitemShut
  {NoStop}%
\bibitem [{\citenamefont {Chen}(2018{\natexlab{c}})}]{Chen:2018hnx}%
  \BibitemOpen
  \bibfield  {author} {\bibinfo {author} {\bibfnamefont {J.-K.}\ \bibnamefont
  {Chen}},\ }\href {\doibase 10.1140/epjc/s10052-018-5718-z} {\bibfield
  {journal} {\bibinfo  {journal} {Eur. Phys. J. C}\ }\textbf {\bibinfo {volume}
  {78}},\ \bibinfo {pages} {235} (\bibinfo {year}
  {2018}{\natexlab{c}})}\BibitemShut {NoStop}%
\bibitem [{\citenamefont {Erlich}\ \emph {et~al.}(2005)\citenamefont {Erlich},
  \citenamefont {Katz}, \citenamefont {Son},\ and\ \citenamefont
  {Stephanov}}]{Erlich:2005qh}%
  \BibitemOpen
  \bibfield  {author} {\bibinfo {author} {\bibfnamefont {J.}~\bibnamefont
  {Erlich}}, \bibinfo {author} {\bibfnamefont {E.}~\bibnamefont {Katz}},
  \bibinfo {author} {\bibfnamefont {D.~T.}\ \bibnamefont {Son}}, \ and\
  \bibinfo {author} {\bibfnamefont {M.~A.}\ \bibnamefont {Stephanov}},\ }\href
  {\doibase 10.1103/PhysRevLett.95.261602} {\bibfield  {journal} {\bibinfo
  {journal} {Phys. Rev. Lett.}\ }\textbf {\bibinfo {volume} {95}},\ \bibinfo
  {pages} {261602} (\bibinfo {year} {2005})},\ \Eprint
  {http://arxiv.org/abs/hep-ph/0501128} {arXiv:hep-ph/0501128} \BibitemShut
  {NoStop}%
\bibitem [{\citenamefont {Braga}\ \emph {et~al.}(2016)\citenamefont {Braga},
  \citenamefont {Martin~Contreras},\ and\ \citenamefont
  {Diles}}]{Braga:2015lck}%
  \BibitemOpen
  \bibfield  {author} {\bibinfo {author} {\bibfnamefont {N.~R.~F.}\
  \bibnamefont {Braga}}, \bibinfo {author} {\bibfnamefont {M.~A.}\ \bibnamefont
  {Martin~Contreras}}, \ and\ \bibinfo {author} {\bibfnamefont
  {S.}~\bibnamefont {Diles}},\ }\href {\doibase 10.1209/0295-5075/115/31002}
  {\bibfield  {journal} {\bibinfo  {journal} {EPL}\ }\textbf {\bibinfo {volume}
  {115}},\ \bibinfo {pages} {31002} (\bibinfo {year} {2016})},\ \Eprint
  {http://arxiv.org/abs/1511.06373} {arXiv:1511.06373 [hep-th]} \BibitemShut
  {NoStop}%
\bibitem [{\citenamefont {Gursoy}\ and\ \citenamefont
  {Kiritsis}(2008)}]{Gursoy:2007cb}%
  \BibitemOpen
  \bibfield  {author} {\bibinfo {author} {\bibfnamefont {U.}~\bibnamefont
  {Gursoy}}\ and\ \bibinfo {author} {\bibfnamefont {E.}~\bibnamefont
  {Kiritsis}},\ }\href {\doibase 10.1088/1126-6708/2008/02/032} {\bibfield
  {journal} {\bibinfo  {journal} {JHEP}\ }\textbf {\bibinfo {volume} {02}},\
  \bibinfo {pages} {032} (\bibinfo {year} {2008})},\ \Eprint
  {http://arxiv.org/abs/0707.1324} {arXiv:0707.1324 [hep-th]} \BibitemShut
  {NoStop}%
\bibitem [{\citenamefont {Cherman}\ \emph {et~al.}(2009)\citenamefont
  {Cherman}, \citenamefont {Cohen},\ and\ \citenamefont
  {Werbos}}]{Cherman:2008eh}%
  \BibitemOpen
  \bibfield  {author} {\bibinfo {author} {\bibfnamefont {A.}~\bibnamefont
  {Cherman}}, \bibinfo {author} {\bibfnamefont {T.~D.}\ \bibnamefont {Cohen}},
  \ and\ \bibinfo {author} {\bibfnamefont {E.~S.}\ \bibnamefont {Werbos}},\
  }\href {\doibase 10.1103/PhysRevC.79.045203} {\bibfield  {journal} {\bibinfo
  {journal} {Phys. Rev. C}\ }\textbf {\bibinfo {volume} {79}},\ \bibinfo
  {pages} {045203} (\bibinfo {year} {2009})},\ \Eprint
  {http://arxiv.org/abs/0804.1096} {arXiv:0804.1096 [hep-ph]} \BibitemShut
  {NoStop}%
\bibitem [{\citenamefont {Boschi-Filho}\ \emph {et~al.}(2013)\citenamefont
  {Boschi-Filho}, \citenamefont {Braga}, \citenamefont {Jugeau},\ and\
  \citenamefont {Torres}}]{Boschi-Filho:2012ijd}%
  \BibitemOpen
  \bibfield  {author} {\bibinfo {author} {\bibfnamefont {H.}~\bibnamefont
  {Boschi-Filho}}, \bibinfo {author} {\bibfnamefont {N.~R.~F.}\ \bibnamefont
  {Braga}}, \bibinfo {author} {\bibfnamefont {F.}~\bibnamefont {Jugeau}}, \
  and\ \bibinfo {author} {\bibfnamefont {M.~A.~C.}\ \bibnamefont {Torres}},\
  }\href {\doibase 10.1140/epjc/s10052-013-2540-5} {\bibfield  {journal}
  {\bibinfo  {journal} {Eur. Phys. J. C}\ }\textbf {\bibinfo {volume} {73}},\
  \bibinfo {pages} {2540} (\bibinfo {year} {2013})},\ \Eprint
  {http://arxiv.org/abs/1208.2291} {arXiv:1208.2291 [hep-th]} \BibitemShut
  {NoStop}%
\bibitem [{\citenamefont {Brambilla}\ \emph {et~al.}(2019)\citenamefont
  {Brambilla}, \citenamefont {Eidelman}, \citenamefont {Hanhart}, \citenamefont
  {Nefediev}, \citenamefont {Shen}, \citenamefont {Thomas}, \citenamefont
  {Vairo},\ and\ \citenamefont {Yuan}}]{Brambilla:2019esw}%
  \BibitemOpen
  \bibfield  {author} {\bibinfo {author} {\bibfnamefont {N.}~\bibnamefont
  {Brambilla}}, \bibinfo {author} {\bibfnamefont {S.}~\bibnamefont {Eidelman}},
  \bibinfo {author} {\bibfnamefont {C.}~\bibnamefont {Hanhart}}, \bibinfo
  {author} {\bibfnamefont {A.}~\bibnamefont {Nefediev}}, \bibinfo {author}
  {\bibfnamefont {C.-P.}\ \bibnamefont {Shen}}, \bibinfo {author}
  {\bibfnamefont {C.~E.}\ \bibnamefont {Thomas}}, \bibinfo {author}
  {\bibfnamefont {A.}~\bibnamefont {Vairo}}, \ and\ \bibinfo {author}
  {\bibfnamefont {C.-Z.}\ \bibnamefont {Yuan}},\ }\href@noop {} {\  (\bibinfo
  {year} {2019})},\ \Eprint {http://arxiv.org/abs/1907.07583} {arXiv:1907.07583
  [hep-ex]} \BibitemShut {NoStop}%
%%CITATION = ARXIV:1907.07583;%%
\bibitem [{\citenamefont {Guo}\ \emph {et~al.}(2018)\citenamefont {Guo},
  \citenamefont {Hanhart}, \citenamefont {Meißner}, \citenamefont {Wang},
  \citenamefont {Zhao},\ and\ \citenamefont {Zou}}]{Guo:2017jvc}%
  \BibitemOpen
  \bibfield  {author} {\bibinfo {author} {\bibfnamefont {F.-K.}\ \bibnamefont
  {Guo}}, \bibinfo {author} {\bibfnamefont {C.}~\bibnamefont {Hanhart}},
  \bibinfo {author} {\bibfnamefont {U.-G.}\ \bibnamefont {Meißner}}, \bibinfo
  {author} {\bibfnamefont {Q.}~\bibnamefont {Wang}}, \bibinfo {author}
  {\bibfnamefont {Q.}~\bibnamefont {Zhao}}, \ and\ \bibinfo {author}
  {\bibfnamefont {B.-S.}\ \bibnamefont {Zou}},\ }\href {\doibase
  10.1103/RevModPhys.90.015004} {\bibfield  {journal} {\bibinfo  {journal}
  {Rev. Mod. Phys.}\ }\textbf {\bibinfo {volume} {90}},\ \bibinfo {pages}
  {015004} (\bibinfo {year} {2018})},\ \Eprint
  {http://arxiv.org/abs/1705.00141} {arXiv:1705.00141 [hep-ph]} \BibitemShut
  {NoStop}%
%%CITATION = ARXIV:1705.00141;%%
\bibitem [{\citenamefont {Lebed}\ \emph {et~al.}(2017)\citenamefont {Lebed},
  \citenamefont {Mitchell},\ and\ \citenamefont {Swanson}}]{Lebed:2016hpi}%
  \BibitemOpen
  \bibfield  {author} {\bibinfo {author} {\bibfnamefont {R.~F.}\ \bibnamefont
  {Lebed}}, \bibinfo {author} {\bibfnamefont {R.~E.}\ \bibnamefont {Mitchell}},
  \ and\ \bibinfo {author} {\bibfnamefont {E.~S.}\ \bibnamefont {Swanson}},\
  }\href {\doibase 10.1016/j.ppnp.2016.11.003} {\bibfield  {journal} {\bibinfo
  {journal} {Prog. Part. Nucl. Phys.}\ }\textbf {\bibinfo {volume} {93}},\
  \bibinfo {pages} {143} (\bibinfo {year} {2017})},\ \Eprint
  {http://arxiv.org/abs/1610.04528} {arXiv:1610.04528 [hep-ph]} \BibitemShut
  {NoStop}%
%%CITATION = ARXIV:1610.04528;%%
\bibitem [{\citenamefont {Liu}\ \emph {et~al.}(2019{\natexlab{a}})\citenamefont
  {Liu}, \citenamefont {Nowak},\ and\ \citenamefont
  {Zahed}}]{PhysRevD.100.126023}%
  \BibitemOpen
  \bibfield  {author} {\bibinfo {author} {\bibfnamefont {Y.}~\bibnamefont
  {Liu}}, \bibinfo {author} {\bibfnamefont {M.~A.}\ \bibnamefont {Nowak}}, \
  and\ \bibinfo {author} {\bibfnamefont {I.}~\bibnamefont {Zahed}},\ }\href
  {\doibase 10.1103/PhysRevD.100.126023} {\bibfield  {journal} {\bibinfo
  {journal} {Phys. Rev. D}\ }\textbf {\bibinfo {volume} {100}},\ \bibinfo
  {pages} {126023} (\bibinfo {year} {2019}{\natexlab{a}})}\BibitemShut
  {NoStop}%
\bibitem [{\citenamefont {Workman}\ \emph {et~al.}(2022)\citenamefont {Workman}
  \emph {et~al.}}]{Workman:2022ynf}%
  \BibitemOpen
  \bibfield  {author} {\bibinfo {author} {\bibfnamefont {R.~L.}\ \bibnamefont
  {Workman}} \emph {et~al.} (\bibinfo {collaboration} {Particle Data Group}),\
  }\href {\doibase 10.1093/ptep/ptac097} {\bibfield  {journal} {\bibinfo
  {journal} {PTEP}\ }\textbf {\bibinfo {volume} {2022}},\ \bibinfo {pages}
  {083C01} (\bibinfo {year} {2022})}\BibitemShut {NoStop}%
\bibitem [{\citenamefont {Hou}\ \emph {et~al.}(2001)\citenamefont {Hou},
  \citenamefont {Luo},\ and\ \citenamefont {Wong}}]{Hou:2001ig}%
  \BibitemOpen
  \bibfield  {author} {\bibinfo {author} {\bibfnamefont {W.-S.}\ \bibnamefont
  {Hou}}, \bibinfo {author} {\bibfnamefont {C.-S.}\ \bibnamefont {Luo}}, \ and\
  \bibinfo {author} {\bibfnamefont {G.-G.}\ \bibnamefont {Wong}},\ }\href
  {\doibase 10.1103/PhysRevD.64.014028} {\bibfield  {journal} {\bibinfo
  {journal} {Phys. Rev.}\ }\textbf {\bibinfo {volume} {D64}},\ \bibinfo {pages}
  {014028} (\bibinfo {year} {2001})},\ \Eprint
  {http://arxiv.org/abs/hep-ph/0101146} {arXiv:hep-ph/0101146 [hep-ph]}
  \BibitemShut {NoStop}%
%%CITATION = HEP-PH/0101146;%%
\bibitem [{\citenamefont {Kruczenski}\ \emph {et~al.}(2003)\citenamefont
  {Kruczenski}, \citenamefont {Mateos}, \citenamefont {Myers},\ and\
  \citenamefont {Winters}}]{Kruczenski:2003be}%
  \BibitemOpen
  \bibfield  {author} {\bibinfo {author} {\bibfnamefont {M.}~\bibnamefont
  {Kruczenski}}, \bibinfo {author} {\bibfnamefont {D.}~\bibnamefont {Mateos}},
  \bibinfo {author} {\bibfnamefont {R.~C.}\ \bibnamefont {Myers}}, \ and\
  \bibinfo {author} {\bibfnamefont {D.~J.}\ \bibnamefont {Winters}},\ }\href
  {\doibase 10.1088/1126-6708/2003/07/049} {\bibfield  {journal} {\bibinfo
  {journal} {JHEP}\ }\textbf {\bibinfo {volume} {07}},\ \bibinfo {pages} {049}
  (\bibinfo {year} {2003})},\ \Eprint {http://arxiv.org/abs/hep-th/0304032}
  {arXiv:hep-th/0304032} \BibitemShut {NoStop}%
\bibitem [{\citenamefont {Jaffe}(2005)}]{Jaffe:2004ph}%
  \BibitemOpen
  \bibfield  {author} {\bibinfo {author} {\bibfnamefont {R.~L.}\ \bibnamefont
  {Jaffe}},\ }\bibfield  {booktitle} {\emph {\bibinfo {booktitle}
  {{Proceedings, 6th International Conference on Hyperons, charm and beauty
  hadrons (BEACH 2004): Chicago, USA, June 27-July 3, 2004}}},\ }\href
  {\doibase 10.1016/j.physrep.2004.11.005} {\bibfield  {journal} {\bibinfo
  {journal} {Phys. Rept.}\ }\textbf {\bibinfo {volume} {409}},\ \bibinfo
  {pages} {1} (\bibinfo {year} {2005})},\ \bibinfo {note} {[,191(2004)]},\
  \Eprint {http://arxiv.org/abs/hep-ph/0409065} {arXiv:hep-ph/0409065 [hep-ph]}
  \BibitemShut {NoStop}%
%%CITATION = HEP-PH/0409065;%%
\bibitem [{\citenamefont {Wang}(2018)}]{Wang:2018ntv}%
  \BibitemOpen
  \bibfield  {author} {\bibinfo {author} {\bibfnamefont {Z.-G.}\ \bibnamefont
  {Wang}},\ }\href {\doibase 10.1140/epjc/s10052-018-6417-5} {\bibfield
  {journal} {\bibinfo  {journal} {Eur. Phys. J.}\ }\textbf {\bibinfo {volume}
  {C78}},\ \bibinfo {pages} {933} (\bibinfo {year} {2018})},\ \Eprint
  {http://arxiv.org/abs/1809.10299} {arXiv:1809.10299 [hep-ph]} \BibitemShut
  {NoStop}%
%%CITATION = ARXIV:1809.10299;%%
\bibitem [{\citenamefont {Liu}\ \emph {et~al.}(2019{\natexlab{b}})\citenamefont
  {Liu}, \citenamefont {Chen}, \citenamefont {Chen}, \citenamefont {Liu},\ and\
  \citenamefont {Zhu}}]{Liu:2019zoy}%
  \BibitemOpen
  \bibfield  {author} {\bibinfo {author} {\bibfnamefont {Y.-R.}\ \bibnamefont
  {Liu}}, \bibinfo {author} {\bibfnamefont {H.-X.}\ \bibnamefont {Chen}},
  \bibinfo {author} {\bibfnamefont {W.}~\bibnamefont {Chen}}, \bibinfo {author}
  {\bibfnamefont {X.}~\bibnamefont {Liu}}, \ and\ \bibinfo {author}
  {\bibfnamefont {S.-L.}\ \bibnamefont {Zhu}},\ }\href {\doibase
  10.1016/j.ppnp.2019.04.003} {\bibfield  {journal} {\bibinfo  {journal} {Prog.
  Part. Nucl. Phys.}\ }\textbf {\bibinfo {volume} {107}},\ \bibinfo {pages}
  {237} (\bibinfo {year} {2019}{\natexlab{b}})},\ \Eprint
  {http://arxiv.org/abs/1903.11976} {arXiv:1903.11976 [hep-ph]} \BibitemShut
  {NoStop}%
%%CITATION = ARXIV:1903.11976;%%
\bibitem [{\citenamefont {Voloshin}(2008)}]{Voloshin:2007dx}%
  \BibitemOpen
  \bibfield  {author} {\bibinfo {author} {\bibfnamefont {M.~B.}\ \bibnamefont
  {Voloshin}},\ }\href {\doibase 10.1016/j.ppnp.2008.02.001} {\bibfield
  {journal} {\bibinfo  {journal} {Prog. Part. Nucl. Phys.}\ }\textbf {\bibinfo
  {volume} {61}},\ \bibinfo {pages} {455} (\bibinfo {year} {2008})},\ \Eprint
  {http://arxiv.org/abs/0711.4556} {arXiv:0711.4556 [hep-ph]} \BibitemShut
  {NoStop}%
%%CITATION = ARXIV:0711.4556;%%
\bibitem [{\citenamefont {Brambilla}\ \emph {et~al.}(2017)\citenamefont
  {Brambilla}, \citenamefont {Shtabovenko}, \citenamefont {Tarrús~Castellà},\
  and\ \citenamefont {Vairo}}]{Brambilla:2017ffe}%
  \BibitemOpen
  \bibfield  {author} {\bibinfo {author} {\bibfnamefont {N.}~\bibnamefont
  {Brambilla}}, \bibinfo {author} {\bibfnamefont {V.}~\bibnamefont
  {Shtabovenko}}, \bibinfo {author} {\bibfnamefont {J.}~\bibnamefont
  {Tarrús~Castellà}}, \ and\ \bibinfo {author} {\bibfnamefont
  {A.}~\bibnamefont {Vairo}},\ }\href {\doibase 10.1103/PhysRevD.95.116004}
  {\bibfield  {journal} {\bibinfo  {journal} {Phys. Rev.}\ }\textbf {\bibinfo
  {volume} {D95}},\ \bibinfo {pages} {116004} (\bibinfo {year} {2017})},\
  \Eprint {http://arxiv.org/abs/1704.03476} {arXiv:1704.03476 [hep-ph]}
  \BibitemShut {NoStop}%
%%CITATION = ARXIV:1704.03476;%%
\bibitem [{\citenamefont {Gleiser}\ and\ \citenamefont
  {Stamatopoulos}(2012)}]{Gleiser:2012tu}%
  \BibitemOpen
  \bibfield  {author} {\bibinfo {author} {\bibfnamefont {M.}~\bibnamefont
  {Gleiser}}\ and\ \bibinfo {author} {\bibfnamefont {N.}~\bibnamefont
  {Stamatopoulos}},\ }\href {\doibase 10.1103/PhysRevD.86.045004} {\bibfield
  {journal} {\bibinfo  {journal} {Phys. Rev. D}\ }\textbf {\bibinfo {volume}
  {86}},\ \bibinfo {pages} {045004} (\bibinfo {year} {2012})},\ \Eprint
  {http://arxiv.org/abs/1205.3061} {arXiv:1205.3061 [hep-th]} \BibitemShut
  {NoStop}%
\bibitem [{\citenamefont {Bernardini}\ \emph {et~al.}(2017)\citenamefont
  {Bernardini}, \citenamefont {Braga},\ and\ \citenamefont
  {da~Rocha}}]{Bernardini:2016qit}%
  \BibitemOpen
  \bibfield  {author} {\bibinfo {author} {\bibfnamefont {A.~E.}\ \bibnamefont
  {Bernardini}}, \bibinfo {author} {\bibfnamefont {N.~R.~F.}\ \bibnamefont
  {Braga}}, \ and\ \bibinfo {author} {\bibfnamefont {R.}~\bibnamefont
  {da~Rocha}},\ }\href {\doibase 10.1016/j.physletb.2016.12.007} {\bibfield
  {journal} {\bibinfo  {journal} {Phys. Lett. B}\ }\textbf {\bibinfo {volume}
  {765}},\ \bibinfo {pages} {81} (\bibinfo {year} {2017})},\ \Eprint
  {http://arxiv.org/abs/1609.01258} {arXiv:1609.01258 [hep-th]} \BibitemShut
  {NoStop}%
\bibitem [{\citenamefont {Braga}\ and\ \citenamefont
  {da~Rocha}(2017)}]{Braga:2016wzx}%
  \BibitemOpen
  \bibfield  {author} {\bibinfo {author} {\bibfnamefont {N.~R.~F.}\
  \bibnamefont {Braga}}\ and\ \bibinfo {author} {\bibfnamefont
  {R.}~\bibnamefont {da~Rocha}},\ }\href {\doibase
  10.1016/j.physletb.2017.02.031} {\bibfield  {journal} {\bibinfo  {journal}
  {Phys. Lett. B}\ }\textbf {\bibinfo {volume} {767}},\ \bibinfo {pages} {386}
  (\bibinfo {year} {2017})},\ \Eprint {http://arxiv.org/abs/1612.03289}
  {arXiv:1612.03289 [hep-th]} \BibitemShut {NoStop}%
\bibitem [{\citenamefont {Bernardini}\ and\ \citenamefont
  {da~Rocha}(2016)}]{Bernardini:2016hvx}%
  \BibitemOpen
  \bibfield  {author} {\bibinfo {author} {\bibfnamefont {A.~E.}\ \bibnamefont
  {Bernardini}}\ and\ \bibinfo {author} {\bibfnamefont {R.}~\bibnamefont
  {da~Rocha}},\ }\href {\doibase 10.1016/j.physletb.2016.09.023} {\bibfield
  {journal} {\bibinfo  {journal} {Phys. Lett. B}\ }\textbf {\bibinfo {volume}
  {762}},\ \bibinfo {pages} {107} (\bibinfo {year} {2016})},\ \Eprint
  {http://arxiv.org/abs/1605.00294} {arXiv:1605.00294 [hep-th]} \BibitemShut
  {NoStop}%
\bibitem [{\citenamefont {Braga}\ \emph {et~al.}(2018)\citenamefont {Braga},
  \citenamefont {Ferreira},\ and\ \citenamefont {Da~Rocha}}]{Braga:2018fyc}%
  \BibitemOpen
  \bibfield  {author} {\bibinfo {author} {\bibfnamefont {N.~R.~F.}\
  \bibnamefont {Braga}}, \bibinfo {author} {\bibfnamefont {L.~F.}\ \bibnamefont
  {Ferreira}}, \ and\ \bibinfo {author} {\bibfnamefont {R.~a.}\ \bibnamefont
  {Da~Rocha}},\ }\href {\doibase 10.1016/j.physletb.2018.10.036} {\bibfield
  {journal} {\bibinfo  {journal} {Phys. Lett. B}\ }\textbf {\bibinfo {volume}
  {787}},\ \bibinfo {pages} {16} (\bibinfo {year} {2018})},\ \Eprint
  {http://arxiv.org/abs/1808.10499} {arXiv:1808.10499 [hep-ph]} \BibitemShut
  {NoStop}%
\bibitem [{\citenamefont {Braga}\ and\ \citenamefont
  {da~Mata}(2020{\natexlab{a}})}]{Braga:2020hhs}%
  \BibitemOpen
  \bibfield  {author} {\bibinfo {author} {\bibfnamefont {N.~R.~F.}\
  \bibnamefont {Braga}}\ and\ \bibinfo {author} {\bibfnamefont
  {R.}~\bibnamefont {da~Mata}},\ }\href {\doibase
  10.1016/j.physletb.2020.135918} {\bibfield  {journal} {\bibinfo  {journal}
  {Phys. Lett. B}\ }\textbf {\bibinfo {volume} {811}},\ \bibinfo {pages}
  {135918} (\bibinfo {year} {2020}{\natexlab{a}})},\ \Eprint
  {http://arxiv.org/abs/2008.10457} {arXiv:2008.10457 [hep-th]} \BibitemShut
  {NoStop}%
\bibitem [{\citenamefont {Braga}\ and\ \citenamefont
  {da~Mata}(2020{\natexlab{b}})}]{Braga:2020myi}%
  \BibitemOpen
  \bibfield  {author} {\bibinfo {author} {\bibfnamefont {N.~R.~F.}\
  \bibnamefont {Braga}}\ and\ \bibinfo {author} {\bibfnamefont
  {R.}~\bibnamefont {da~Mata}},\ }\href {\doibase 10.1103/PhysRevD.101.105016}
  {\bibfield  {journal} {\bibinfo  {journal} {Phys. Rev. D}\ }\textbf {\bibinfo
  {volume} {101}},\ \bibinfo {pages} {105016} (\bibinfo {year}
  {2020}{\natexlab{b}})},\ \Eprint {http://arxiv.org/abs/2002.09413}
  {arXiv:2002.09413 [hep-th]} \BibitemShut {NoStop}%
\bibitem [{\citenamefont {Braga}\ and\ \citenamefont
  {Junqueira}(2021)}]{Braga:2020opg}%
  \BibitemOpen
  \bibfield  {author} {\bibinfo {author} {\bibfnamefont {N.~R.~F.}\
  \bibnamefont {Braga}}\ and\ \bibinfo {author} {\bibfnamefont {O.~C.}\
  \bibnamefont {Junqueira}},\ }\href {\doibase 10.1016/j.physletb.2021.136082}
  {\bibfield  {journal} {\bibinfo  {journal} {Phys. Lett. B}\ }\textbf
  {\bibinfo {volume} {814}},\ \bibinfo {pages} {136082} (\bibinfo {year}
  {2021})},\ \Eprint {http://arxiv.org/abs/2010.00714} {arXiv:2010.00714
  [hep-th]} \BibitemShut {NoStop}%
\bibitem [{\citenamefont {Martin~Contreras}\ \emph {et~al.}(2022)\citenamefont
  {Martin~Contreras}, \citenamefont {Vega},\ and\ \citenamefont
  {Diles}}]{MartinContreras:2022lxl}%
  \BibitemOpen
  \bibfield  {author} {\bibinfo {author} {\bibfnamefont {M.~A.}\ \bibnamefont
  {Martin~Contreras}}, \bibinfo {author} {\bibfnamefont {A.}~\bibnamefont
  {Vega}}, \ and\ \bibinfo {author} {\bibfnamefont {S.}~\bibnamefont {Diles}},\
  }\href {\doibase 10.1016/j.physletb.2022.137551} {\bibfield  {journal}
  {\bibinfo  {journal} {Phys. Lett. B}\ }\textbf {\bibinfo {volume} {835}},\
  \bibinfo {pages} {137551} (\bibinfo {year} {2022})},\ \Eprint
  {http://arxiv.org/abs/2206.01834} {arXiv:2206.01834 [hep-ph]} \BibitemShut
  {NoStop}%
\bibitem [{\citenamefont {Polchinski}\ and\ \citenamefont
  {Strassler}(2002)}]{Polchinski:2001tt}%
  \BibitemOpen
  \bibfield  {author} {\bibinfo {author} {\bibfnamefont {J.}~\bibnamefont
  {Polchinski}}\ and\ \bibinfo {author} {\bibfnamefont {M.~J.}\ \bibnamefont
  {Strassler}},\ }\href {\doibase 10.1103/PhysRevLett.88.031601} {\bibfield
  {journal} {\bibinfo  {journal} {Phys. Rev. Lett.}\ }\textbf {\bibinfo
  {volume} {88}},\ \bibinfo {pages} {031601} (\bibinfo {year} {2002})},\
  \Eprint {http://arxiv.org/abs/hep-th/0109174} {arXiv:hep-th/0109174}
  \BibitemShut {NoStop}%
\bibitem [{\citenamefont {Balasubramanian}\ \emph {et~al.}(1999)\citenamefont
  {Balasubramanian}, \citenamefont {Kraus}, \citenamefont {Lawrence},\ and\
  \citenamefont {Trivedi}}]{Balasubramanian:1998de}%
  \BibitemOpen
  \bibfield  {author} {\bibinfo {author} {\bibfnamefont {V.}~\bibnamefont
  {Balasubramanian}}, \bibinfo {author} {\bibfnamefont {P.}~\bibnamefont
  {Kraus}}, \bibinfo {author} {\bibfnamefont {A.~E.}\ \bibnamefont {Lawrence}},
  \ and\ \bibinfo {author} {\bibfnamefont {S.~P.}\ \bibnamefont {Trivedi}},\
  }\href {\doibase 10.1103/PhysRevD.59.104021} {\bibfield  {journal} {\bibinfo
  {journal} {Phys. Rev. D}\ }\textbf {\bibinfo {volume} {59}},\ \bibinfo
  {pages} {104021} (\bibinfo {year} {1999})},\ \Eprint
  {http://arxiv.org/abs/hep-th/9808017} {arXiv:hep-th/9808017} \BibitemShut
  {NoStop}%
\bibitem [{\citenamefont {Hamilton}\ \emph {et~al.}(2006)\citenamefont
  {Hamilton}, \citenamefont {Kabat}, \citenamefont {Lifschytz},\ and\
  \citenamefont {Lowe}}]{Hamilton:2006az}%
  \BibitemOpen
  \bibfield  {author} {\bibinfo {author} {\bibfnamefont {A.}~\bibnamefont
  {Hamilton}}, \bibinfo {author} {\bibfnamefont {D.~N.}\ \bibnamefont {Kabat}},
  \bibinfo {author} {\bibfnamefont {G.}~\bibnamefont {Lifschytz}}, \ and\
  \bibinfo {author} {\bibfnamefont {D.~A.}\ \bibnamefont {Lowe}},\ }\href
  {\doibase 10.1103/PhysRevD.74.066009} {\bibfield  {journal} {\bibinfo
  {journal} {Phys. Rev. D}\ }\textbf {\bibinfo {volume} {74}},\ \bibinfo
  {pages} {066009} (\bibinfo {year} {2006})},\ \Eprint
  {http://arxiv.org/abs/hep-th/0606141} {arXiv:hep-th/0606141} \BibitemShut
  {NoStop}%
\bibitem [{\citenamefont {Rehren}(2000)}]{Rehren:2000tp}%
  \BibitemOpen
  \bibfield  {author} {\bibinfo {author} {\bibfnamefont {K.-H.}\ \bibnamefont
  {Rehren}},\ }\href {\doibase 10.1016/S0370-2693(00)01168-0} {\bibfield
  {journal} {\bibinfo  {journal} {Phys. Lett. B}\ }\textbf {\bibinfo {volume}
  {493}},\ \bibinfo {pages} {383} (\bibinfo {year} {2000})},\ \Eprint
  {http://arxiv.org/abs/hep-th/0003120} {arXiv:hep-th/0003120} \BibitemShut
  {NoStop}%
\bibitem [{\citenamefont {Banks}\ \emph {et~al.}(1998)\citenamefont {Banks},
  \citenamefont {Douglas}, \citenamefont {Horowitz},\ and\ \citenamefont
  {Martinec}}]{Banks:1998dd}%
  \BibitemOpen
  \bibfield  {author} {\bibinfo {author} {\bibfnamefont {T.}~\bibnamefont
  {Banks}}, \bibinfo {author} {\bibfnamefont {M.~R.}\ \bibnamefont {Douglas}},
  \bibinfo {author} {\bibfnamefont {G.~T.}\ \bibnamefont {Horowitz}}, \ and\
  \bibinfo {author} {\bibfnamefont {E.~J.}\ \bibnamefont {Martinec}},\
  }\href@noop {} {\  (\bibinfo {year} {1998})},\ \Eprint
  {http://arxiv.org/abs/hep-th/9808016} {arXiv:hep-th/9808016} \BibitemShut
  {NoStop}%
\bibitem [{\citenamefont {Bena}(2000)}]{Bena:1999jv}%
  \BibitemOpen
  \bibfield  {author} {\bibinfo {author} {\bibfnamefont {I.}~\bibnamefont
  {Bena}},\ }\href {\doibase 10.1103/PhysRevD.62.066007} {\bibfield  {journal}
  {\bibinfo  {journal} {Phys. Rev. D}\ }\textbf {\bibinfo {volume} {62}},\
  \bibinfo {pages} {066007} (\bibinfo {year} {2000})},\ \Eprint
  {http://arxiv.org/abs/hep-th/9905186} {arXiv:hep-th/9905186} \BibitemShut
  {NoStop}%
\bibitem [{\citenamefont {Braga}\ \emph {et~al.}(2017)\citenamefont {Braga},
  \citenamefont {Ferreira},\ and\ \citenamefont {Vega}}]{Braga:2017bml}%
  \BibitemOpen
  \bibfield  {author} {\bibinfo {author} {\bibfnamefont {N.~R.~F.}\
  \bibnamefont {Braga}}, \bibinfo {author} {\bibfnamefont {L.~F.}\ \bibnamefont
  {Ferreira}}, \ and\ \bibinfo {author} {\bibfnamefont {A.}~\bibnamefont
  {Vega}},\ }\href {\doibase 10.1016/j.physletb.2017.10.013} {\bibfield
  {journal} {\bibinfo  {journal} {Phys. Lett. B}\ }\textbf {\bibinfo {volume}
  {774}},\ \bibinfo {pages} {476} (\bibinfo {year} {2017})},\ \Eprint
  {http://arxiv.org/abs/1709.05326} {arXiv:1709.05326 [hep-ph]} \BibitemShut
  {NoStop}%
\bibitem [{\citenamefont {Martin~Contreras}\ \emph {et~al.}(2021)\citenamefont
  {Martin~Contreras}, \citenamefont {Diles},\ and\ \citenamefont
  {Vega}}]{MartinContreras:2021bis}%
  \BibitemOpen
  \bibfield  {author} {\bibinfo {author} {\bibfnamefont {M.~A.}\ \bibnamefont
  {Martin~Contreras}}, \bibinfo {author} {\bibfnamefont {S.}~\bibnamefont
  {Diles}}, \ and\ \bibinfo {author} {\bibfnamefont {A.}~\bibnamefont {Vega}},\
  }\href {\doibase 10.1103/PhysRevD.103.086008} {\bibfield  {journal} {\bibinfo
   {journal} {Phys. Rev. D}\ }\textbf {\bibinfo {volume} {103}},\ \bibinfo
  {pages} {086008} (\bibinfo {year} {2021})},\ \Eprint
  {http://arxiv.org/abs/2101.06212} {arXiv:2101.06212 [hep-ph]} \BibitemShut
  {NoStop}%
\bibitem [{\citenamefont {Karapetyan}(2022)}]{Karapetyan:2021crv}%
  \BibitemOpen
  \bibfield  {author} {\bibinfo {author} {\bibfnamefont {G.}~\bibnamefont
  {Karapetyan}},\ }\href {\doibase 10.1140/epjp/s13360-022-02736-1} {\bibfield
  {journal} {\bibinfo  {journal} {Eur. Phys. J. Plus}\ }\textbf {\bibinfo
  {volume} {137}},\ \bibinfo {pages} {590} (\bibinfo {year} {2022})},\ \Eprint
  {http://arxiv.org/abs/2112.11359} {arXiv:2112.11359 [nucl-th]} \BibitemShut
  {NoStop}%
\bibitem [{\citenamefont {Barreto}\ and\ \citenamefont
  {da~Rocha}(2022)}]{Barreto:2022ohl}%
  \BibitemOpen
  \bibfield  {author} {\bibinfo {author} {\bibfnamefont {W.}~\bibnamefont
  {Barreto}}\ and\ \bibinfo {author} {\bibfnamefont {R.}~\bibnamefont
  {da~Rocha}},\ }\href {\doibase 10.1103/PhysRevD.105.064049} {\bibfield
  {journal} {\bibinfo  {journal} {Phys. Rev. D}\ }\textbf {\bibinfo {volume}
  {105}},\ \bibinfo {pages} {064049} (\bibinfo {year} {2022})},\ \Eprint
  {http://arxiv.org/abs/2201.08324} {arXiv:2201.08324 [hep-th]} \BibitemShut
  {NoStop}%
\bibitem [{\citenamefont {Martin~Contreras}\ \emph {et~al.}(2023)\citenamefont
  {Martin~Contreras}, \citenamefont {Vega},\ and\ \citenamefont
  {Diles}}]{MartinContreras:2023eft}%
  \BibitemOpen
  \bibfield  {author} {\bibinfo {author} {\bibfnamefont {M.~A.}\ \bibnamefont
  {Martin~Contreras}}, \bibinfo {author} {\bibfnamefont {A.}~\bibnamefont
  {Vega}}, \ and\ \bibinfo {author} {\bibfnamefont {S.}~\bibnamefont {Diles}},\
  }\href@noop {} {\  (\bibinfo {year} {2023})},\ \Eprint
  {http://arxiv.org/abs/2308.16007} {arXiv:2308.16007 [hep-ph]} \BibitemShut
  {NoStop}%
\end{thebibliography}%
\bibliographystyle{apsrev4-1}
\end{document}